\PassOptionsToPackage{desactivate}{linenoaa}
\documentclass{aa}  
\usepackage{graphicx}
\usepackage{multirow}
\usepackage{array}
\usepackage{booktabs}
\usepackage{float} 
\usepackage{caption} 

\usepackage[colorlinks=true,     linkcolor=blue, citecolor=blue, filecolor=blue, urlcolor=blue]{hyperref}

\usepackage{txfonts}
\usepackage{natbib}
\begin{document} 

   \title{The Atmospheric Composition of Sub-Neptune K2-18\,b and Implications for its Formation}

   \author{Gareb Fern\'andez-Rodr\'iguez\inst{1,2},
          Giuseppe Morello\inst{3,4}, 
          Jonathan C. Tan\inst{2,5},
          Enric Pall\'e\inst{1,6},
          Mark R. Swain\inst{7},
          Efthymios Poultourtzidis\inst{2,8},
          Alfredo Biagini\inst{4},
          Quentin Changeat\inst{9},
          Chengzi Jiang\inst{1,6},
          Francisco J. Pozuelos\inst{3},
          Pedro J. Amado\inst{3}
          }

         \institute{Departamento de Astrof\'isica, Universidad de La Laguna (ULL), 38206 La Laguna, Tenerife, Spain
         \and
         Dept. of Space, Earth $\&$ Environment, Chalmers University of 
         Technology, Gothenburg, Sweden
         \and
         Instituto de Astrof\'isica de Andaluc\'ia (IAA-CSIC), Glorieta de la Astronom\'ia s/n, 18008 Granada, Spain
         \and
         INAF- Palermo Astronomical Observatory, Piazza del Parlamento, 1, 90134 Palermo, Italy
         \and
         Dept. of Astronomy $\&$ Virginia Institute for Theoretical Astronomy, 
         University of Virginia, Charlottesville, VA, USA
         \and
         Instituto de Astrof\'isica de Canarias (IAC), 38205 La Laguna, Tenerife, Spain
         \and 
         California Institute of Technology, NASA Jet Propulsion Laboratory, 4800 Oak Grove Dr, La Cañada Flintridge, CA 91011, USA
         \and
         Department of Physics, Aristotle University of Thessaloniki, University Campus, Thessaloniki, 54124, Greece
         \and
         Kapteyn Institute, University of Groningen, 9747 AD Groningen, NL
         \\
         }

   \date{Received MONTH XX, XXXX; accepted MONTH XX, XXXX}

  \abstract
   {Unlocking the atmospheres of sub-Neptune planets is one of the revolutionary accomplishments of JWST. However such observations require complex data analysis methodologies, which have a strong impact on the derived conclusions. Here, we present an independent reanalysis of the JWST transmission spectrum of the temperate sub-Neptune K2-18\,b, to assess the robustness of previously claimed atmospheric detections, explore the planet's possible parameter space, and implications for its formation. The NIRISS/SOSS and NIRSpec/G395H observations were reduced using a combination of public and customized pipelines. We produced a total of 12 different versions of the transmission spectrum by varying key steps: spectral binning, limb-darkening treatment, and the application of a novel correction for an occulted stellar spot, as well as error inflation and instrumental offsets. We then performed atmospheric retrievals using \texttt{TauREx 3}, comparing models of varying complexity. We robustly detect $\mathrm{CH_4}$ at >4$\sigma$ significance across all reduction and retrieval setups. The evidence for $\mathrm{CO_2}$ is weaker and highly model-dependent, with a typical significance of $\sim2\sigma$. The tentative detection of dimethyl sulphide (DMS), reported in previous studies, vanishes in our most comprehensive retrieval models. We find that correcting the stellar spot in the NIRISS transit is a critical step, introducing a uniform offset that primarily drives the inference of a lighter, lower mean molecular weight atmosphere. Furthermore, the assumed complexity of the retrieval model itself introduces significant biases; including more molecules systematically increases the retrieved $\mathrm{CH_4}$ abundance and atmospheric mean molecular weight, even for species without spectral features. The data are consistent with a hydrogen-rich, i.e., primordial, atmosphere with an elevated O abundance and an even more elevated C abundance, leading to a C/O ratio significantly greater than solar. We show that the physical properties of the K2-18 system planets, i.e., innermost planet K2-18~c, and K2-18~b are consistent with those expected by the {\it in situ} formation theory of Inside-Out Planet Formation (IOPF). Furthermore, these properties predict the temperature of K2-18~b at time of formation was $\gtrsim500\:$K, i.e., much warmer than the current equilibrium temperature and just interior to the carbon ``soot'' line, where an elevated C/O ratio of a primordial atmosphere is expected to be inherited from the protoplanetary disk.}

    \titlerunning{Atmospheric Composition and Formation Constraints for K2-18\,b}
    \authorrunning{G. Fernández-Rodríguez, G. Morello et al.}
   
   \maketitle

\section{Introduction}

K2-18\,b is a temperate sub-Neptune orbiting an M-dwarf star located at 38 pc in the constellation Leo \citep{Montet2015}. With a mass of 8.63$\pm$1.35\,$M_{\oplus}$ \citep{Cloutier2019} and a radius of 2.610$\pm$0.087\,$R_{\oplus}$ \citep{Benneke2019}, the planet lies in a degenerate region of parameter space concerning its interior composition. K2-18\,b could host a water-rich core with a modest H$_2$-He envelope, or a predominantly rocky interior with a thicker primordial atmosphere \citep{LuquePalle2022,Rogers2023}. Intriguingly, the planet orbits within the habitable zone of its host star, receiving a stellar irradiation comparable to that of Earth. This has motivated the proposal of a ``Hycean world'' scenario, in which K2-18\,b might sustain liquid water beneath a hydrogen-rich atmosphere \citep{Madhusudhan2021}.

Spectroscopic observations with HST/WFC3, prior to the launch of JWST, revealed a prominent absorption feature at 1.4\,$\mu$m. It was initially reported as a $\gtrsim$3$\sigma$ detection of water vapour \citep{Benneke2019,Tsiaras2019}. Based on self-consistent 1D radiative equilibrium models, \cite{Blain2021} and \cite{Bezard2022} argued that the the 1.4\,$\mu$m feature was more likely due to methane. \cite{Barclay2021} suggested unocculted stellar spots as an alternative explanation. Recently, JWST solved this dilemma by obtaining a high-precision transmission spectrum of K2-18\,b over 0.9--5.2\,$\mu$m, combining NIRISS/SOSS and NIRSpec/G395H observations. The first analysis of these observations reported CH$_4$ ($\sim$5$\sigma$), CO$_2$ ($\sim$3$\sigma$), and tentative dimethyl sulphide ((CH$_3$)$_2$S or DMS) detections \citep{Madhusudhan2023}. Along with the non-detection of H$_2$O, CO, and NH$_3$, these findings were initially interpreted as indicative of a Hycean scenario for K2-18\,b \citep{Hu2021,Madhusudhan2023}. However, follow-up studies noted that they are consistent with a gas-rich envelope over a magma ocean \citep{Shorttle2024,Wogan2024}. An independent reanalysis of the same JWST data confirmed the CH$_4$ detection, but raised doubts about the CO$_2$ or DMS detections, suggesting that K2-18\,b is most likely an oxygen-poor mini-Neptune without a water ocean \citep{Schmidt2025}. While this manuscript was in preparation, \cite{Hu2025} reported new JWST observations of K2-18\,b covering 1.7--5.2\,$\mu$m, reaffirming the detection of CO$_2$ at 3.7$\sigma$ and, in their interpretation, lending further support to the Hycean world hypothesis.

The tentative detection of DMS in K2-18\,b has sparked interest due to its potential as a biosignature gas \citep{Seager2013}, though an abiotic origin cannot be ruled out \citep{Reed2024}. A 6--12\,$\mu$m transmission spectrum obtained with JWST/MIRI LRS was initially reported as further evidence for DMS or dimethyl disulphide (DMDS) \citep{Madhusudhan2025}, though this interpretation has been widely debated \citep{Luque2025,Stevenson2025,Taylor2025,Welbanks2025}.

In this work, we present an independent reanalysis of JWST observations of K2-18\,b using NIRSpec/G395H and NIRISS/SOSS data (GO 2722, PI: Madhusudhan). We carefully examine how choices in data reduction, light-curve modelling, and retrieval configurations affect the inferred atmospheric properties. In particular, at the data processing level, we investigate the effects of spectral binning at the light-curve or transmission-spectrum stage, treatment of stellar limb-darkening and spots, instrumental offsets, and error-bar rescaling. At the modelling stage, we investigate the effect the number of included molecules, temperature–pressure profiles, and the presence of clouds and haze have in the retrievals. An additional result of our investigation is a newly developed semi-empirical method for correcting occulted stellar spots in spectral light curves, that has been proven effective. This investigation underscores the importance of repeated JWST observations to assess the robustness of molecular detections and highlights the value of future missions such as Ariel \citep{Tinetti2018}, whose simultaneous broad-wavelength coverage will help mitigate instrumental offsets and astrophysical variability. Beyond K2-18\,b, our analysis also aims to provide guidelines for interpreting JWST observations of sub-Neptune atmospheres from single-transit datasets, such as TOI-421\,b \citep{Davenport2025} or GJ 9827\,d \citep{Piaulet2024}. Then, in the final part of our paper we examine the implications of our derived atmospheric properties for K2-18~b for its formation, concluding that evidence points to a H$_2$-He dominated atmosphere with a high C/O ratio that is consistent with a scenario of {\it in situ} formation of a volatile-poor planet.

\section{Observations}

Two full transits of K2-18\,b were observed with the JWST as part of GO Program 2722 (PI: N. Madhusudhan). The first transit was observed from January 20 to January 21 2023 between 18:37:38 and 01:11:32 UTC for a total of $5.3$ hours. This observation was made using NIRSpec (\citealp{Ferruit2012}, \citealp{Jakobsen2022}, \citealp{Birkmann2022}) with the G395H grating, Bright Object Time Series mode (BOTS), F290LP filter, SUB2048 subarray, and the NRSRAPID readout pattern. The resulting spectrum was measured with the NRS1 and NRS2 detectors, spanning wavelengths of $2.8-3.72\, \mathrm{\mu m}$ and $3.83-5.17\, \mathrm{\mu m}$ each, with a detector gap between $3.72$ and $3.83 \, \mathrm{\mu m}$. 

The second observation took place during June 1 2023 from 13:49:20 to 19:36:05 UTC for a total of $4.9$ hours. It used the NIRISS instrument (\citealp{Doyon2012,Doyon2023}) in the Single Object Slitless Spectroscopy, SOSS, mode (\citealp{Albert2023}), with the GR700XD grism, CLEAR filter, SUBSTRIP256 subarray, and NISRAPID readout pattern, resulting in a wavelength coverage of $0.85-2.83\, \mathrm{\mu m}$ for the first spectral order and $0.6-1.1\, \mathrm{\mu m}$ for the second spectral order. 
\section{Data Reduction}

We reduced the observations starting from the raw data making use of a combination of publicly available pipelines and a self-developed pipeline. The first 3 stages of the JWST Calibration pipeline (\texttt{jwst}, \citealp{Bushouse2023}), which includes detector-level corrections, wavelength calibration and spectral trace extraction were carried out using standard reduction pipelines. After that we used a custom-built pipeline for stages 4-6, which includes white light curve and spectral light curve fitting and extraction of the transmission spectrum. This approach allows us to test the effects of different decisions made during the reduction process on the final transmission spectrum.

\subsection{NIRISS SOSS}
The first three stages of NIRISS data reduction were performed using \texttt{exoTEDRF} \citep{Feinstein2023, Radica2023,Radica2024}. Our reduction follows the typical reduction process for sub-Neptunes and super-Earths \citep{Lim2023,Cadieux2024,Benneke2024}. Stage 1 includes flagging of saturated pixels, subtraction of the super-bias, correction of the reference pixel position, and subtraction of the dark current. In the jump detection step we used a threshold of 7$\sigma$ for the time-domain outlier rejection algorithm \citep{Radica2024a}, similarly to \cite{Ahrer2025}, although we use a more conservative value. A scaled model group-level background is temporally subtracted from the data in order to better remove the $1/f$ noise which affects all JWST instruments \citep{Schlawin2020}. After removal of the aforementioned noise, the background model is re-added and the ramp linearity is fitted. Background photons are recorded simultaneously with target photons, therefore the background inherits the same non-linearity effects and should be corrected after linearity calibration \citep{Feinstein2023,Radica2023}. Our implementation of Stage 2 includes the default parameters, photometric calibration, and wavelength mapping, although we skip the flat-fielding step. In Stage 3, we perform spectral extraction using a box extraction method, assuming that the trace width remains constant, with an aperture of 30 pixels. This approach has been extensively used in prior JWST analyses \citep{Davenport2025}. In general the two spectral orders in NIRISS SOSS overlap, therefore there will be some spectral contamination. More complex extraction algorithms exist that account for this effect, like the ATOCA algorithm, however, these effects can be neglected in relative flux measurements \citep{Darveau2022,Radica2022}.

\begin{table}[htbp]
\centering
\small
\begin{tabular}{l c c}
\toprule
Parameter & Prior & Posterior \\
\midrule
\multicolumn{3}{l}{\textbf{NIRISS}} \\
$R_\mathrm{p}/R_\mathrm{s}$ & $\mathcal{U}(0,\,1)$ & $0.05400^{+0.00016}_{-0.00016}$ \\
$T_0^a$ & $\mathcal{N}(60096.729374,\,0.000035)$ & $60096.729369^{+0.000032}_{-0.000031}$ \\
$P$ [d] & $\mathcal{N}(32.939623,\,0.0000012)$ & $32.9396230^{+0.0000012}_{-0.0000013}$ \\
$\tau _0$ [s] & $\mathcal{U}(0,\,50000)$ & $8855.4^{+10.9}_{-10.1}$ \\
$b$ & $\mathcal{U}(0,\,1)$ & $0.611^{+0.016}_{-0.018}$ \\
$\phi_\mathrm{spot}$ [$^{\circ}$] & $\mathcal{U}(-80,\,-10)$ & $-31.3^{+6.3}_{-7.0}$ \\
$\lambda_\mathrm{spot}$ [$^{\circ}$] & $\mathcal{U}(-70,\,-15)$ & $-47.9^{+2.5}_{-3.8}$ \\
$f_\mathrm{disk}/f_\mathrm{spot}$ & $\mathcal{U}(0,\,1)$ & $0.816^{+0.026}_{-0.035}$ \\
$r_\mathrm{disk}/r_\mathrm{spot}$ & $\mathcal{U}(0.01,\,0.35)$ & $0.259^{+0.058}_{-0.057}$ \\

\midrule
\multicolumn{3}{l}{\textbf{NIRSpec NRS1}} \\
$R_\mathrm{p}/R_\mathrm{s}$ & $\mathcal{U}(0,\,1)$ & $0.054254^{+0.000093}_{-0.000094}$ \\
$T_0^a$ & $\mathcal{N}(59964.969453,\,0.000035)$ & $59964.969452^{+0.000026}_{-0.000026}$ \\
$P$ [d] & $\mathcal{N}(32.939623,\,0.0000012)$ & $32.9396230^{+0.0000012}_{-0.0000012}$ \\
$\tau _0$ [s] & $\mathcal{U}(0,\,50000)$ & $8872.4^{+6.8}_{-6.9}$ \\
$b$ & $\mathcal{U}(0,\,1)$ & $0.612^{+0.010}_{-0.010}$ \\

\midrule
\multicolumn{3}{l}{\textbf{NIRSpec NRS2}} \\
$R_\mathrm{p}/R_\mathrm{s}$ & $\mathcal{U}(0,\,1)$ & $0.05402^{+0.00013}_{-0.00013}$ \\

\bottomrule
\end{tabular}
\caption{Priors and posteriors for each instrument. Orbital parameters in the case of NIRSpec white light curve fits are shared.}
\tablefoot{$a$: Format is BJD TDB - 2400000.5}
\label{tab:priors_white_lc}
\end{table}

\begin{figure*}[ht]
    \centering
    \includegraphics[width=\textwidth]{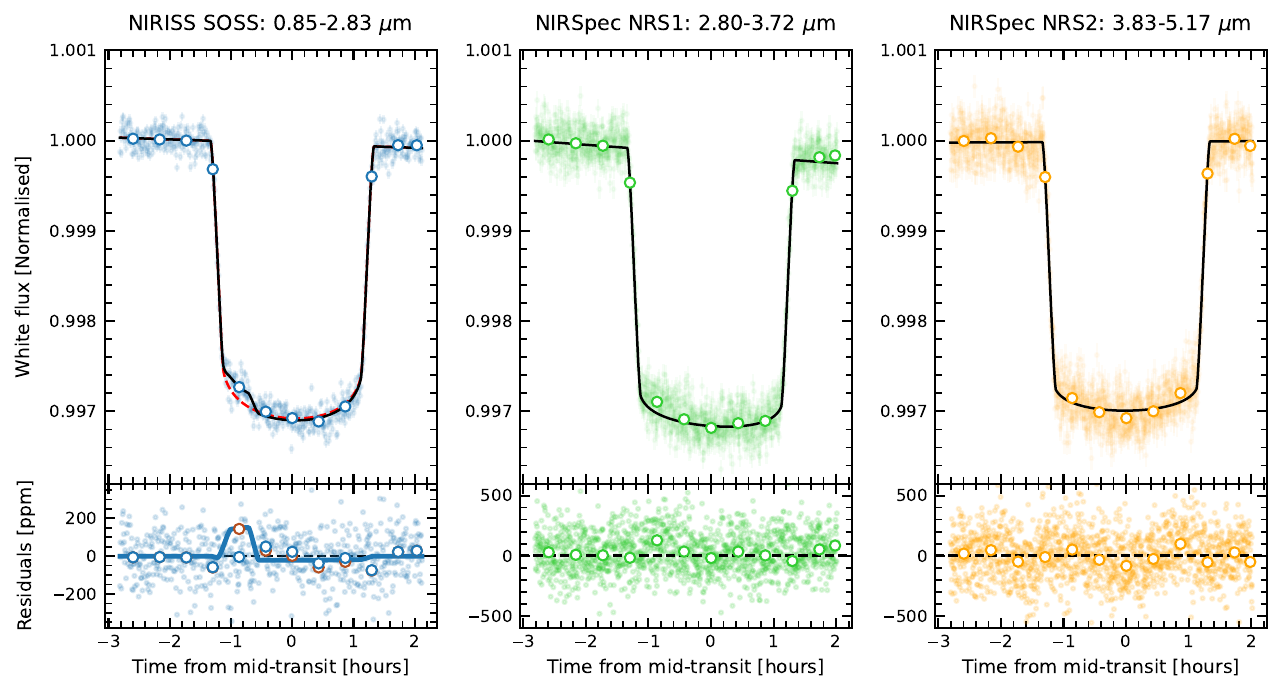}
    \caption{White light curves for NIRISS NIRSpec NRS1 and NIRSpec NRS2 with best fit models. The blue line in the left-most panel residuals (NIRISS) indicates the resulting profile from the subtraction of the spot and spotless models. The red dashed line is the spotless model. The coloured points are 10 minute binned light curves.}
    \label{fig:white_light_curves}
\end{figure*}

\subsection{NIRSpec G395H}
Both detectors (NRS1 and NRS2) were reduced independently with identical reduction parameters making use of the \texttt{Eureka!} pipeline \citep{Bell2022}. In Stage 1, we flagged saturated pixels, subtracted the super-bias, corrected the reference pixel positions, and applied the ramp linearity correction to obtain count rates and subtract the dark current. Curved traces were also masked. We increased the jump detection threshold to 10$\sigma$, in line with the ERS team recommendations \citep{Alderson2023,Rustamkulov2023}. The $1/f$ noise (vertical striping pattern) is also corrected in this step by removing the column-wise median of pixels located outside the trace. Stage 2 of \texttt{Eureka!} is essentially a wrapper of the corresponding \texttt{jwst} pipeline stage \citep{Bushouse2023}. During this stage, wavelength calibration is applied. We used the default parameters for this step, but excluded photometric calibrations from the reduction process \citep{Alderson2023}. In Stage 3, we performed the spectral extraction. The region of interest was set from rows 0 to 31 along the spatial axis and columns 500 to 2042 along the spectral axis. We adopted a 10$\sigma$ threshold for outlier rejection during background subtraction. The curvature of the trace was corrected by shifting the position of the pixel columns. This is done such that peak of the count distribution in the detector, which we assume to be Gaussian, lies in the centre of window, in our case row 15. The spectrum was then extracted using a 7-pixel half-width aperture centred on the trace. The background was estimated using the region 14 pixels away from the trace, well outside our extraction box.

\begin{figure*}[ht]
    \centering
    \includegraphics[width=\textwidth]{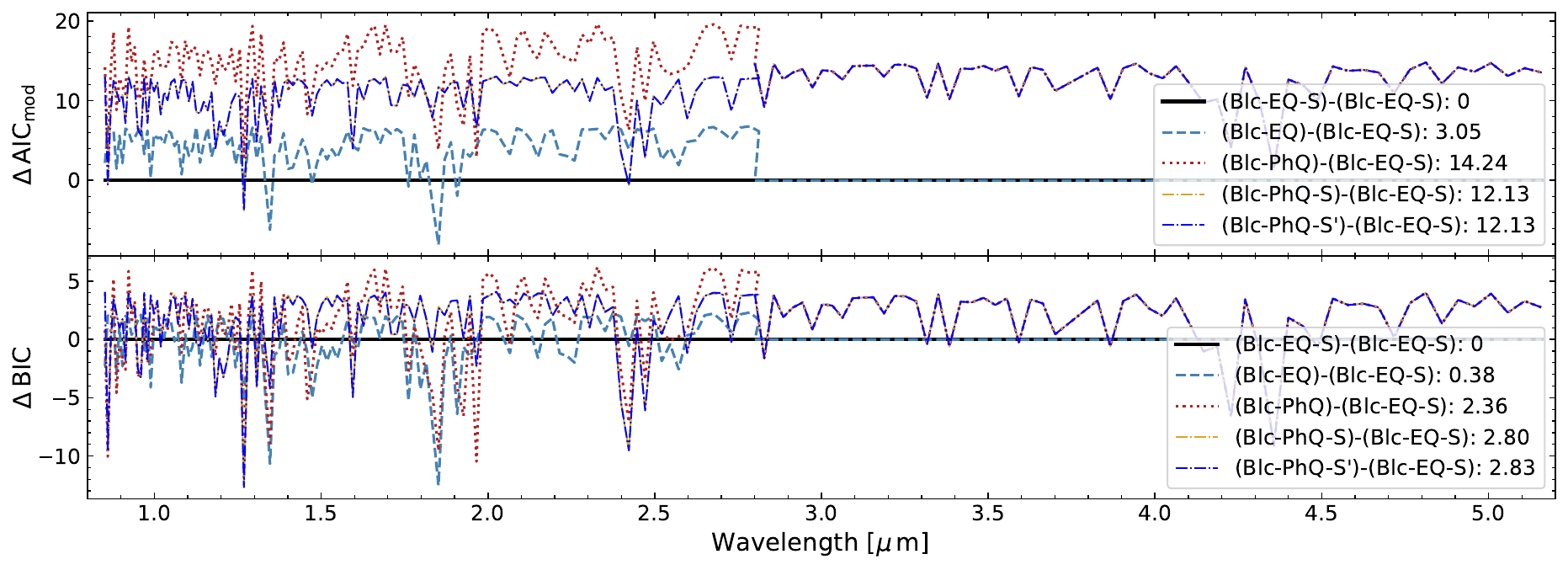}
    \caption{Comparison between different reduction modified AICs in the top panel and BICs in the lower panel (all reduction names and their characteristics are described in Section \ref{sec:spec_lc_fits}) taking the Blc-EQ-S (binned before fitting light curves - empirical limb darkening - with spot correction) reduction as baseline. The median of the modified AIC or BIC is shown in the legend. For those cases where the NIRSpec data is shared between reductions, only the median for the NIRISS data is shown. Although there are peaks below, the majority of the points lie above the 0 $\Delta \mathrm{\,AIC_{mod}}$ and $\Delta \mathrm{\,BIC}$ lines.}
    \label{fig:aics}
\end{figure*}

\section{Light-curve Analysis}
\subsection{White-light curves}
For the white light curve analysis (Stage 4 of data reduction) we used a custom built pipeline. The white light curves were obtained by summing the flux within the spectral extraction aperture for each frame. In this way, we extracted three light curves, one from each detector. These time series were fitted using a transit model with a polynomial baseline (order $\leq$2) to account for instrumental systematics. The basic free transit parameters included the planet-to-star radius ratio $(R_p/R_s)$, orbital period ($P$), mid-transit time $(T_0)$, impact parameter ($b$), and a transit duration parameter assuming a point-like planet ($\tau_0$). The selection of the latter two parameters follows the recommendations of \citet{Morello2018}, where formulae relating them to orbital parameters are provided. Limb-darkening coefficients were fixed to theoretical values corresponding to a four-parameter law \citep{Claret2020}, computed using the \texttt{ExoTETHyS} package \citep{Morello2020JOSS,Morello2020A}. These coefficients were derived from the spherical one-dimensional (1D) \texttt{PHOENIX} stellar atmosphere model \citep{Husser2013}, adopting stellar parameters of $T_{\rm eff} = 3496\:$K and $\log (g / \mathrm{[cm\,s^{-2}]}) = 4.858$ \citep{Crossfield2016}. 
Additionally, we included an error-scaling factor $\left(\gamma\right)$ to rescale the uncertainties and ensure that the reduced $\chi^2$ is close to unity.

The NIRISS SOSS white light curve integrates all flux within the $0.85$–$2.83\,\mu$m range. This data set includes a spot crossing event near the end of transit ingress, as reported by \cite{Madhusudhan2023}. We adopted a specialized software to simulate the planetary transits in front of a limb-darkened stellar disc with spots \citep{Cracchiolo2021,Biagini2024}. The stellar disk is divided radially and azimuthally into 1000 steps each, resulting in a grid of one million elements. The flux contribution from each element is computed at each time step during the transit, based on the planet's position. A single spot is included in the fit, described by four free parameters: latitude ($\phi_{\mathrm{spot}}$), longitude ($\lambda_{\mathrm{spot}}$), the intensity ratio ($f_{\mathrm{spot}} / f_{\mathrm{disk}}$) at the spot's centre, and the spot radius ratio ($r_{\mathrm{spot}} / r_{\mathrm{disk}}$). Model selection based on the Bayesian Information Criterion (BIC; \citealp{Schwarz1978}) indicates a first-order polynomial baseline with two free parameters ($r_0$, $r_1$) as the preferred choice for the NIRISS SOSS data as opposed to constant or second-order polynomial models, which yielded higher BIC values by $\sim$8.

From NIRSpec, the NRS1 and NRS2 light curves integrate $2.80-3.72\,\mu$m and $3.83-5.17\,\mu$m, respectively. In this case, we assumed a simpler transit model generated with \texttt{PYLIGHTCURVE} \citep{Tsiaras2016}. Our choice is motivated by the lack of statistical evidence for spots, also in the original study of \cite{Madhusudhan2023}. The NRS1 and NRS2 light curves were fitted simultaneously with shared orbital parameters ($P$, $T_0$, $b$, and $\mathrm{\tau _0}$), but independent radius ratios $(R_p/R_s)$, due to its possible wavelength-dependence. We find that the derived orbital parameter values are within uncertainties to those from the NIRISS fit, see Table \ref{tab:priors_white_lc}. Similarly to NIRISS a linear ramp was preferred, as constant or quadratic alternatives resulted in higher BIC values by 5 and 7, respectively.

A summary of the most relevant priors and retrieved parameters can be found in Table \ref{tab:priors_white_lc}. The priors for the orbital period, mid transit times, were taken from \cite{Madhusudhan2023},
while the others are uniform. The resulting fitted light curves can be seen in Figure \ref{fig:white_light_curves}.

\subsection{Spectroscopic light curves}
\label{sec:spec_lc_fits}

Spectroscopic light curves were extracted from the same apertures as the white ones, using either single-pixel columns or flux integrated over wavelength bins. For all spectroscopic fits, the orbital parameters ($P$, $T_0$, $b$ and $\tau_0$) were fixed to the values derived from the white light curves, while $R_p/R_s$, the baseline ramp parameters and error scaling factor were left free.
We ran several approaches to obtain the transmission spectrum, testing the impact of varying the following procedures:
\begin{itemize}
\item \textbf{Spectral binning.} The transmission spectrum was derived from binned light curves (Blc) at a resolution of $R\sim100$, following recommendations from the JWST Transiting Exoplanet Community Early Release Science team \citep{Carter2024}, or from light curves at the native pixel resolution. For comparison, we also derived the binned transmission spectrum (Bts) from that obtained at the native pixel resolution. Previous studies have suggested that $R\sim100$ represents the lowest resolution at which JWST transmission spectra retain most of the information, providing an optimal compromise between computational efficiency and reliability of atmospheric retrievals \citep{Davey2025};
\item \textbf{Limb-darkening.} We fitted the binned light curves allowing for empirical quadratic limb-darkening coefficients (EQ), or fixing them to the corresponding \texttt{PHOENIX}-model values (PhQ) computed with \texttt{ExoTETHyS}. For the single-column light curves, we always fixed the limb-darkening coefficients to \texttt{PHOENIX}-model values, considering both the quadratic (PhQ) and four-coefficient (PhC4) laws.
\item \textbf{Occulted spot.} We introduce a new method to account for the spot-crossing feature in spectroscopic light-curve fitting. The spot signal is obtained as the difference between the spotted transit model from the white light-curve fit and the corresponding spotless model, created by setting $r_\mathrm{spot}/r_\mathrm{disk} = 0$. We then assume that the spot signal shape is consistent across wavelengths, but with a varying amplitude. Accordingly, the spectroscopic light curves are fitted as a linear combination of a spotless transit model and the spot signal ($S$). As a more advanced approach, we computed an ad hoc spot signal for each spectral bin by adjusting the limb-darkening coefficients to theoretical values ($S^\prime$). We also ran tests without spot correction to quantify its importance.
\end{itemize}

\begin{figure*}[ht]
    \centering
    \includegraphics[width=\textwidth]{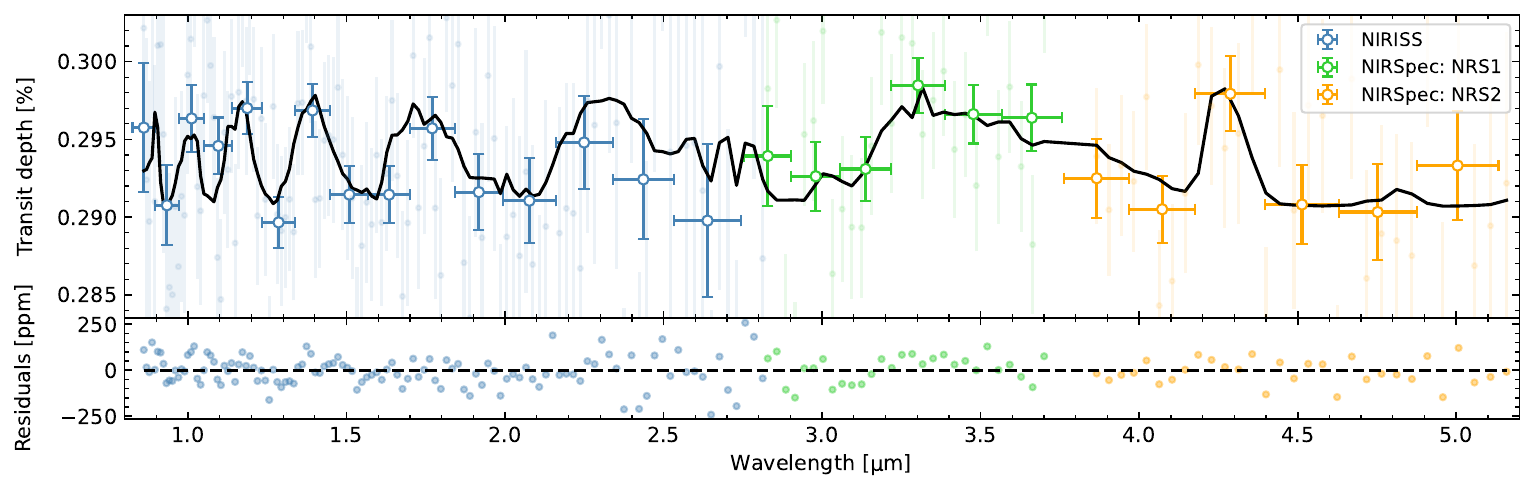}
    \caption{Fiducial reduction and retrieval which includes: spot correction, 2 coefficient empirical limb darkening law, light curves binned to a resolution of 100 before the fit (Blc-EQ-S). The data (lighter shaded points) are binned to a resolution of $R\sim15$ (darker shaded points) to highlight the spectral features. The bottom panel shows the residuals, at $R\sim100$, which is the resolution at which the retrieval was done.}
    \label{fig:spectrum}
\end{figure*}

\subsection{Fiducial transmission spectrum}
\label{sec:fid_spec_lc}

We performed extensive retrieval analyses on all extracted spectra to evaluate how the inferred atmospheric properties depend on choices made during data reduction, light-curve fitting, and retrieval configuration. Not all extractions are equally reliable. Previous studies have shown that JWST high-resolution modes are particularly sensitive to analysis details such as limb-darkening, likely due to non-Gaussian parameter correlations in low signal-to-noise (S/N) light curve fits \citep{Carter2024}. For this reason, we consider the Blc extractions more robust, as the binning increases the effective S/N of the fitted light curves. Within the Blc subset, those accounting for the NIRISS spot-crossing event are regarded as the most trustworthy. The use of empirical limb-darkening coefficients further minimizes model dependence and reduces potential biases from theoretical prescriptions. Accordingly, we adopt the Blc-EQ-S spectrum as our fiducial reference.

Moreover this is shown quantitatively by comparing the modified Akaike information criterion (AIC, \citealp{Akaike1974}) and the BIC at each wavelength bin and deriving an average across all wavelengths, as shown in Figure \ref{fig:aics}. Both methods confirm that indeed the Blc-EQ-S reduction is moderately preferred.

\section{Retrievals}
\label{sec:retrievals_method}

%\subsection{TauREx}
We conducted a suite of free-chemistry retrievals on multiple variants of the combined transmission spectrum using \texttt{TauREx} 3.1 \citep{Waldmann2015, Al-Refaie2021}. The planetary atmosphere was modelled numerically with 100 layers, evenly spaced on a logarithmic pressure scale from 10$^{-7}$ to 10$^{2}$ bar. Our retrieval setups included absorption from selected molecular species, collision-induced absorption by H$_2$–H$_2$ and H$_2$-He \citep{Abel2011,Abel2012,Fletcher2018}, Mie scattering following the formalism of \cite{Lee2013}, and an optically thick cloud deck defined by its top pressure. The temperature–pressure profile was parametrized using the formulation of \cite{Guillot2010}, or a simpler isothermal model for comparison. For the former, we assumed an internal temperature of 60\,K, which is well within the range predicted by previous studies \citep{Blain2021,Madhusudhan2023}.

We considered absorption from the most prominent molecular species expected in mini-Neptune atmospheres, including H$_2$O \citep{Polyansky2018}, CO \citep{Li2015}, CO$_2$ \citep{Yurchenko2020}, CH$_4$ \citep{Yurchenko2017}, NH$_3$ \citep{Coles2019}, HCN \citep{Barber2014}, H$_2$S \citep{Azzam2016}, and SO$_2$ \citep{Underwood2016} (resolutions$~5\times10^4$). Additionally, we considered potential biosignatures tested by \cite{Madhusudhan2023}, though adopting newer opacity data sources for some species, such as OCS \citep{Owens2024}, CH$_3$Cl \citep{Owens2018}, N$_2$O \citep{Yurchenko2024}, CS$_2$ \citep{Karlovets2021}, and (CH$_3$)$_2$S or DMS. Note that for DMS and CS$_2$, we extrapolated the absorption cross sections provided directly by HITRAN \citep{Sharpe2004,Kochanov2019,Gordon2022} at 1\,bar and 298\,K to other temperature and pressure conditions, following the method of \cite{Madhusudhan2021,Madhusudhan2023}. Retrievals were also carried out with selected subsets of molecules to assess the statistical significance of different contributions and evaluate how the inferred physical parameters and chemical abundances were affected by the species considered.

The atmospheric parameter space was sampled using the nested sampling algorithm \texttt{PyMultiNest} \citep{Feroz2009,Buchner2014}, with 3000 live points to ensure a thorough exploration. We adopted broad (log-)uniform parameter priors, which are largely uninformative. In particular, the molecular mixing ratios were allowed to vary across 17 orders of magnitude, ranging from negligible quantities to as much as 99\% of the total. Along with the posterior distributions of atmospheric parameters, \texttt{PyMultiNest} provides the Bayesian evidence for each retrieval model, which is particularly useful for comparing nested models \citep{Trotta2008,Tsiaras2018}. For instance, the detection significance of a specific molecule can be inferred from the difference in log-evidence between otherwise identical models, differing only by the inclusion of that molecule, a method widely used in recent studies (e.g., \citealp{Guilluy2021,Madhusudhan2023,Powell2024}).

\begin{table*}
\caption{Selected parameters from retrievals on the fiducial transmission spectrum (Blc-EQ-S).}            % title of Table
\label{tab:comp_retr_fid}      % is used to refer this table in the text
\centering                                      % used for centering table
\begin{tabular}{ccccccccc}          % centered columns (4 columns)
\hline\hline
Molecules & $\chi_0^2$ & $\log{\mathrm{ev}}$ & $\log{(\mathrm{CO}_2)}$ & $\log{(\mathrm{CH}_4)}$ & $T_{\mathrm{10 \, mbar}}$ (K) & $MMW$ & $\log{(P_\mathrm{cloud}/\mathrm{bar})}$ \\
\hline
%\multicolumn{10}{c}{Baseline setup: Guillot T-P profile, Cloud deck \& Mie scattering}\\
Flat\tablefootmark{a} & 1.26 & 1440.86 & -- & -- & -- & -- & -- \\
CO$_2$ & 1.34 & $-0.34$ & N & -- \\
CH$_4$ & 1.19 & $+7.66$ & -- & $-1.78_{-0.76}^{+0.64}$ (4.32$\sigma$) & $124_{-38}^{+44}$ & $2.53_{-0.19}^{+0.75}$ & $-1.30_{-0.60}^{+1.24}$ \\
DMS & 1.28 & $+0.61$ & -- & -- & \\
\textbf{CO$_2$+CH$_4$}\tablefootmark{b} & \textbf{1.17} & $\textbf{+8.62}$ & $-\textbf{3.94}_{-4.20}^{+1.68}$ \textbf{(2.00$\sigma$)} & $-\textbf{1.73}_{-0.73}^{+0.66}$ \textbf{(4.62$\sigma$)} & $\textbf{138}_{-39}^{+48}$ & $\textbf{2.64}_{-0.28}^{+1.16}$ & $\textbf{-1.14}_{-0.59}^{+1.35}$ \\
CO$_2$+CH$_4$+DMS & 1.18 & $+8.52$ & $-4.05_{-4.86}^{+1.83}$ & $-1.65_{-0.77}^{+0.65}$ & $126_{-38}^{+47}$ & $2.71_{-0.34}^{+1.31}$ & $-1.02_{-0.75}^{+1.53}$ \\
All\tablefootmark{c} & 1.25 & $+7.49$ & $-3.63_{-5.13}^{+1.75}$ (N) & $-0.96_{-0.57}^{+0.45}$ (4.16$\sigma$) & $161_{-54}^{+90}$ & 4.32$_{-1.45}^{+3.02}$ & $-0.88_{-1.10}^{+1.59}$ \\
\hline
\end{tabular}
\tablefoot{Retrieval setups including Guillot T-P profiles, an  optically-thick cloud deck, and Mie scattering. The significance of detection for molecular species are reported in parenthesis, N denotes a non-detection. \tablefoottext{a}{Forcedly flat spectrum obtained by imposing the highest cloud deck, but adjusting the radius to better match the data}; \tablefoottext{b}{Simplified retrieval setup, including molecules with at least a 2$\sigma$ detection}; \tablefoottext{c}{Comprehensive retrieval setup, including the 13 molecular species considered in this study.} }
\end{table*}

\begin{table*}
\caption{Selected parameters from retrievals on different spectral extractions, including the simplified and comprehensive retrieval setups.}            % title of Table
\label{tab:retr_all}      % is used to refer this table in the text
\centering                                      % used for centering table
\begin{tabular}{cccccccc}          % centered columns (4 columns)
\hline\hline
Spectrum & Molecules & $\chi_0^2$ & $\log{\mathrm{ev}}$ & $\log{(\mathrm{CO}_2)}$ & $\log{(\mathrm{CH}_4)}$ & $T_{\mathrm{10 \, mbar}}$ (K) & $MMW$ \\
\hline
\multirow{2}{*}{\textbf{Blc-EQ-S}} & \textbf{CO$_2$+CH$_4$} & \textbf{1.17} & \textbf{1449.47} & $\textbf{-3.94}_{-4.20}^{+1.68}$ \textbf{(2.00$\sigma$)} & $\textbf{-1.73}_{-0.73}^{+0.66}$ \textbf{(4.62$\sigma$)} & $\textbf{138}_{-39}^{+48}$ & $\textbf{2.64}_{-0.28}^{+1.16}$ \\
 & All & 1.25 & 1448.35 & $-3.63_{-5.13}^{+1.75}$ (N) & $-0.96_{-0.57}^{+0.45}$ (4.16$\sigma$) & $161_{-54}^{+90}$ & 4.32$_{-1.45}^{+3.02}$ \\
\hline
\multirow{2}{*}{Blc-PhQ-S} & CH$_4$ & 1.31 & 1450.59 & -- & $-1.95_{-0.75}^{+0.67}$ (4.56$\sigma$) & $121_{-35}^{+44}$ & $2.46_{-0.13}^{+0.56}$ \\
 & All & 1.39 & 1449.88 & $-4.76_{-5.30}^{+2.56}$ (N) & $-0.95_{-0.60}^{+0.45}$ (4.06$\sigma$) & $146_{-48}^{+89}$ & $4.26_{-1.44}^{+3.05}$ \\
\hline
\multirow{2}{*}{Blc-PhQ-S$'$} & CH$_4$ & 1.30 & 1450.94 & -- & $-1.94_{-0.76}^{+0.68}$ (4.56$\sigma$) & $123_{-37}^{+43}$ & $2.46_{-0.13}^{+0.59}$ \\
 & All & 1.39 & 1450.20 & $-4.98_{-5.24}^{+2.74}$ (N) & $-0.95_{-0.59}^{+0.46}$ (4.01$\sigma$) & $148_{-49}^{+96}$ & $4.25_{-1.44}^{+3.20}$ \\
\hline
\multirow{2}{*}{Blc-EQ} & CO$_2$+CH$_4$ & 1.19 & 1446.63 & $-3.29_{-3.44}^{+1.73}$ (1.99$\sigma$) & $-1.84_{-0.79}^{+0.93}$ (4.52$\sigma$) & $155_{-46}^{+67}$ & $2.63_{-0.28}^{+3.06}$ \\
 & All & 1.27 & 1446.09 & $-3.00_{-5.58}^{+1.56}$ (N) & $-0.80_{-0.59}^{+0.41}$ (3.75$\sigma$) & $195_{-74}^{+118}$ & $5.63_{-2.42}^{+3.74}$ \\
\hline
\multirow{2}{*}{Blc-PhQ} & CO$_2$+CH$_4$ & 1.36 & 1446.61 & $-1.74_{-4.56}^{+0.81}$ (2.30$\sigma$) & $-1.00_{-1.79}^{+0.64}$ (4.96$\sigma$) & $213_{-79}^{+155}$ & $6.73_{-4.40}^{+5.15}$ \\
 & All & 1.45 & 1447.18 & $-2.17_{-5.46}^{+1.07}$ (N) & $-0.53_{-0.51}^{+0.30}$ (3.59$\sigma$) & $283_{-132}^{+175}$ & $8.83_{-4.27}^{+3.92}$ \\
\hline
\multirow{2}{*}{Bts-PhQ} & CO$_2$+CH$_4$+DMS\tablefootmark{a} & 1.50 & 1443.52 & $-5.54_{-6.07}^{+4.45}$ (3.89$\sigma$) & $-1.57_{-0.89}^{+0.67}$ (5.94$\sigma$) & $134_{-37}^{+80}$ & $2.77_{-0.41}^{+5.31}$ \\
 & All & 1.58 & 1444.07 & $-1.18_{-1.60}^{+0.32}$ (2.33$\sigma$) & $-0.64_{-0.44}^{+0.33}$ (3.83$\sigma$) & $328_{-142}^{+184}$ & $10.14_{-3.39}^{+3.21}$ \\
\hline
\multirow{2}{*}{Bts-PhC4} & CO$_2$+CH$_4$+DMS\tablefootmark{b} & 1.50 & 1443.63 & $-6.79_{-5.23}^{+3.96}$ (3.00$\sigma$) & $-1.80_{-0.78}^{+0.62}$ (5.83$\sigma$) & $125_{-32}^{+42}$ & $2.58_{-0.23}^{+0.84}$ \\
 & All & 1.58 & 1444.01 & $-1.21_{-2.11}^{+0.34}$ (2.17$\sigma$) & $-0.62_{-0.45}^{+0.32}$ (3.85$\sigma$) & $321_{-141}^{+187}$ & $10.20_{-3.59}^{+3.21}$ \\
\hline \hline
\multirow{2}{*}{Bts-Mad23} & CO$_2$+CH$_4$ & 1.56 & 1423.53 & $-1.17_{-1.96}^{+0.24}$ (3.14$\sigma$) & $-1.03_{-1.36}^{+0.43}$ (6.13$\sigma$) & $206_{-63}^{+87}$ & $7.23_{-4.81}^{+2.63}$ \\
 & All & 1.66 & 1423.74 & $-1.40_{-1.44}^{+0.35}$ (2.13$\sigma$) & $-0.70_{-0.42}^{+0.30}$ (3.80$\sigma$) & $248_{-95}^{+144}$ & $7.68_{-2.69}^{+2.62}$ \\
\hline
\multirow{2}{*}{Orig-Mad23} & CO$_2$+CH$_4$+DMS\tablefootmark{c} & 1.08 & 28820.93 & $-1.38_{-0.87}^{+0.32}$ (3.64$\sigma$) & $-1.02_{-0.63}^{+0.41}$ (6.24$\sigma$) & $204_{-73}^{+97}$ & $5.94_{-2.55}^{+2.78}$ \\
 & All & 1.09 & 28820.72 & $-1.44_{-0.57}^{+0.31}$ (2.90$\sigma$) & $-0.84_{-0.40}^{+0.34}$ (4.19$\sigma$) & $254_{-94}^{+129}$ & $6.48_{-1.83}^{+2.35}$ \\
\hline
\end{tabular}
\tablefoot{\tablefoottext{a}{$\log{(\mathrm{DMS})}=-4.60_{-1.70}^{+0.95}$ (2.13$\sigma$)}; \tablefoottext{b}{$\log{(\mathrm{DMS})}=-4.46_{-0.81}^{+0.78}$ (2.31$\sigma$)}; \tablefoottext{c}{$\log{(\mathrm{DMS})}=-6.29_{-5.60}^{+2.06}$ (2.25$\sigma$)}.\\
Blc: transmission spectrum derived from binned light curves at R$\sim$100; 
Bts: binned transmission spectrum obtained from the native pixel resolution; 
EQ, PhQ, PhC4: light-curve fits using empirical quadratic, \texttt{PHOENIX} quadratic, or \texttt{PHOENIX} four-coefficient limb-darkening laws, respectively; 
S, S$'$: spot correction adopting either a single spot signal with a free scaling factor for each wavelength bin, or a wavelength-dependent signal accounting for limb darkening.}
\end{table*}

\begin{figure*}[ht]
    \centering
    \includegraphics[width=\textwidth]{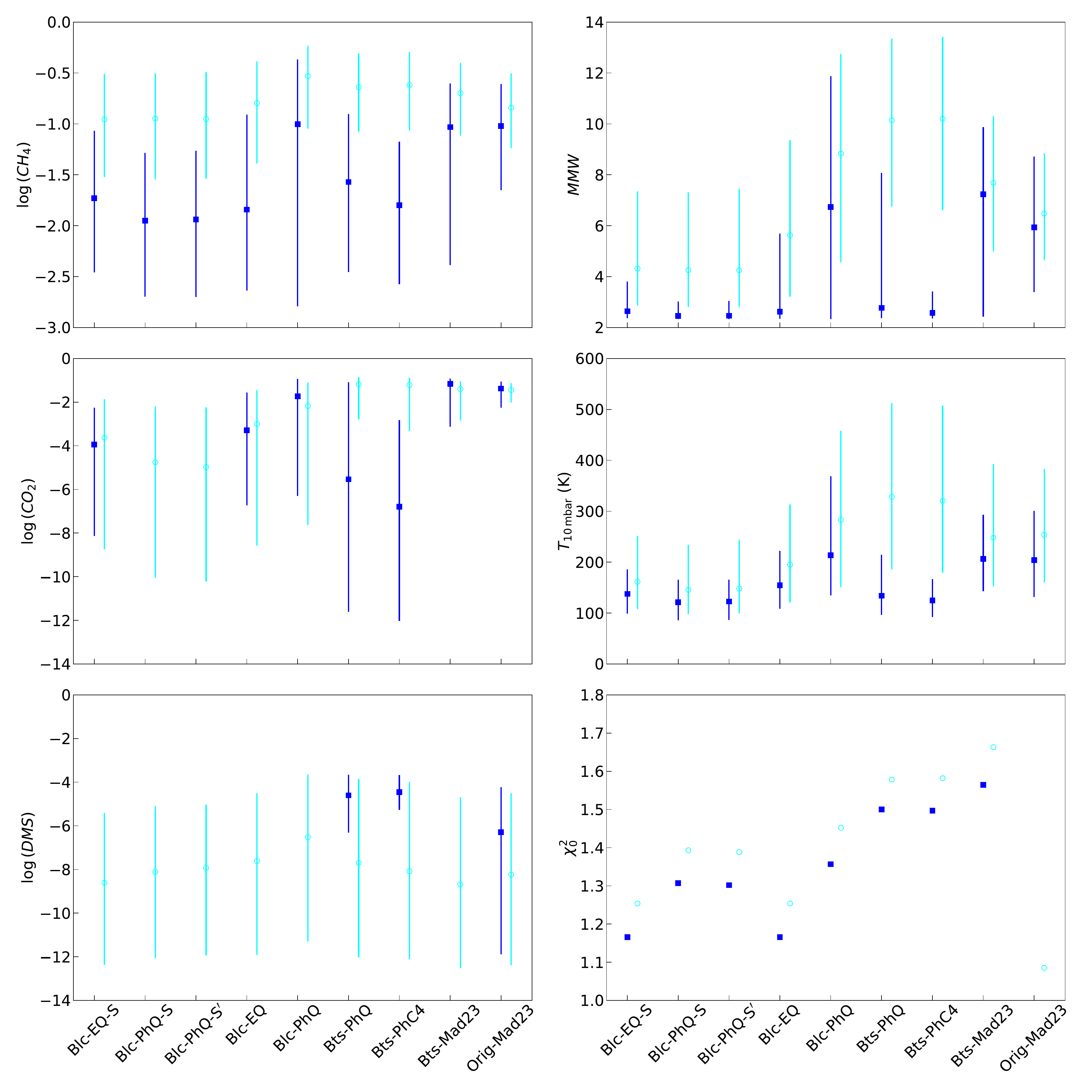}
    \caption{Left: Main molecular abundances obtained from different spectral extractions, using simplified (blue) and comprehensive (cyan) retrieval setups. Note that CO$_2$ was not included in the Blc-PhQ-S and Blc-PhQ-S$'$ simplified retrievals, as its significance was below 2$\sigma$. Similarly, DMS was not included in the majority of simplified retrievals. Right: Retrieved mean molecular weights, temperature at a pressure of 10 mbar, and reduced chi-squared values.}
    \label{fig:main_retr_comp}
\end{figure*}

\begin{table}
\caption{95th-percentile upper limits for the volume mixing ratios of undetected species.}            % title of Table
\label{tab:ref_retr_upperlims}      % is used to refer this table in the text
\centering                                      % used for centering table
\begin{tabular}{cc}          % centered columns (4 columns)
\hline\hline
Molecule & 95$^{th}$ Upper Limit \\
\hline
H$_2$O & <-2.29 / -2.83 \\
CO & <-2.80 / -4.03 \\
NH$_3$ & <-4.41 / -4.96 \\
HCN & <-2.54 / -3.22 \\
H$_2$S & <-2.85 / -3.36 \\
SO$_2$ & <-3.20 / -3.90 \\
DMS & <-4.39 / -4.98 \\
OCS & <-4.61 / -5.77 \\
CH$_3$Cl & <-3.72 / -4.84 \\
N$_2$O & <-3.14 / -3.49 \\
CS$_2$ & <-5.32 / -5.79 \\
\hline
\end{tabular}
\tablefoot{Derived from the Blc-EQ-S spectrum, adopting the comprehensive or simplified retrieval setups, respectively.}
\end{table}

\section{Results}

\subsection{Retrievals from the Fiducial Transmission Spectrum}
\label{sec:retr_fid}

The atmospheric retrievals on our fiducial transmission spectrum confirm CH$_4$ and CO$_2$ as the most likely absorbers, although their absolute abundances and detection significances vary depending on the retrieval setup (see Tables \ref{tab:comp_retr_fid} and \ref{tab:comp_retr_fid2}).

When including all molecular species in an H$_2$-He-dominated atmosphere, a cloud deck, Mie scattering, and assuming a Guillot temperature–pressure profile, we obtain $\log{(\mathrm{CH}_4)}=-0.96_{-0.57}^{+0.45}$ and $\log{(\mathrm{CO}_2)}=-3.63_{-5.13}^{+1.75}$. Notably, the posterior for CO$_2$ is significantly skewed, favouring higher abundances with a peak around $\log{(\mathrm{CO}_2)} \sim -1.88$, but with a long tail extending toward much lower values. Indeed, Bayesian model comparison using a leave-one-out approach indicates that only CH$_4$ is detected with strong statistical significance (4.16$\sigma$). Retrievals excluding CO$_2$ yield comparable Bayesian evidence, suggesting no robust detection of CO$_2$.
For other species, we report only broad 95th-percentile upper limits, such as $\log{(\mathrm{H}_2\mathrm{O})} < -2.29$ and $\log{(\mathrm{NH}_3)} < -4.41$ (see Table \ref{tab:ref_retr_upperlims} for the full list).

Given the lack of evidence for molecules beyond CH$_4$ and CO$_2$, we explored simplified retrievals including one or both of these species, with or without an additional molecule. A model with only CH$_4$ and CO$_2$ yields nearly the highest Bayesian evidence, with additional molecules offering negligible or no improvement. The corresponding abundances are $\log{(\mathrm{CH}_4)} = -1.73_{-0.75}^{+0.66}$ and $\log{(\mathrm{CO}_2)} = -3.94_{-4.20}^{+1.68}$. This solution is only marginally favoured over a pure CH$_4$ scenario with $\log{(\mathrm{CH}_4)} = -1.78_{-0.76}^{+0.64}$, corresponding to a $\sim$2$\sigma$ detection of CO$_2$. Across retrieval setups, the inferred abundances are consistent within 1$\sigma$, though CH$_4$ in particular tends to shift toward higher values when including more molecules. This higher CH$_4$ abundance, together with the high upper limits for certain species, leads to a systematic increase of the mean molecular weight, ranging from $MMW = 2.53_{-0.19}^{+0.75}$ (CH$4$ only) up to $MMW = 4.32_{-1.45}^{+3.02}$ (all molecules).

The retrieved temperature-pressure profiles are nearly isothermal above the 0.01\,bar level, with median posterior temperatures ranging from $T_{\mathrm{10 \, mbar}} \sim  124-161\mathrm{\,K}$ across configurations, and 1$\sigma$ credible intervals spanning $86-251\mathrm{\,K}$. In our reference setup including CO$_2$ and CH$_4$, we obtain $T_{\mathrm{10 , mbar}} = 138_{-39}^{+48}\mathrm{\,K}$.
All retrievals provide consistent hints of a deeper cloud layer, with the preferred setup indicating a cloud top pressure of $\log{(P_\mathrm{cloud}/\mathrm{bar})} = -1.14_{-0.59}^{+1.35}$. The Mie scattering parameters remain poorly constrained.

We also tested retrieval variants assuming an isothermal profile and/or excluding Mie scattering and cloud opacity (see Table \ref{tab:comp_retr_fid2}). These simpler setups yield comparable Bayesian evidences, indicating that neither a thick cloud deck nor Mie scattering is statistically required by the data. The retrieved molecular abundances and temperatures are consistent across setups well within 1$\sigma$.

\subsection{Retrievals from Alternative Spectral Extractions}
\label{sec:retr_other}

All our data processing methods yield visually similar transmission spectra for K2-18\,b, with most points agreeing within 1$\sigma$ (see Appendix \ref{app:transpec_comp}). Nonetheless, even these modest differences can notably affect the inferred atmospheric properties.
We applied the same suite of retrieval setups to each version of the transmission spectrum and ultimately report the ``simplified solution,'' including only molecules with a $\gtrsim2\sigma$ detection, alongside the ``most comprehensive'' model. Key results are summarized in Table \ref{tab:retr_all}  and shown in Figure \ref{fig:main_retr_comp}.

In all cases, CH$_4$ is the most abundant species and is always detected at $>3\sigma$. However, its absolute abundance varies substantially across spectral extractions and retrieval setups, with 1$\sigma$ intervals spanning $\log{(\mathrm{CH}_4)}\in [ -2.79, -0.23]$. The comprehensive retrievals yield systematically higher CH$_4$ abundances than the simplified chemical setups, though they remain consistent within 1$\sigma$.
CO$_2$ is likely the second most abundant molecule, but not always detected with statistical significance. Consequently, the corresponding 1$\sigma$ intervals span an extremely broad range, $\log{(\mathrm{CO}_2)}\in [ -12,-0.93]$.

Retrieval parameters from the Blc spectra agree within fractions of a sigma, particularly when applying spot correction or empirical limb-darkening. The Blc spectra without spot correction exhibit a bimodal solution, with peaks corresponding to H$_2$–He and high–MMW scenarios. Applying the spot correction suppresses the high–MMW peak, particularly in the simplified chemical setup.
The Bts spectra indicate a light or heavy atmosphere, depending on whether the simplified or comprehensive setup is used. These are the only spectra suggesting a $>3\sigma$ detection of CO$_2$ and showing hints of DMS ($\gtrsim2\sigma$) in the optimal retrievals.

\begin{figure*}[ht]
    \centering
    \includegraphics[width=\textwidth]{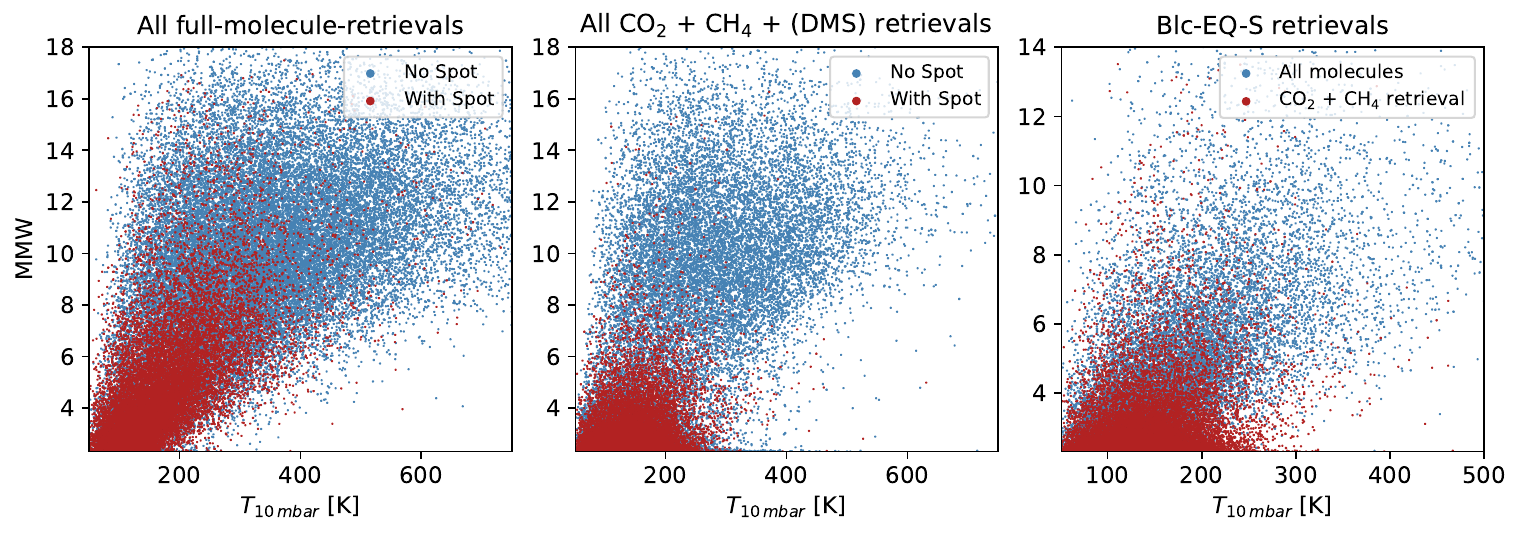}
    \caption{Posterior distribution correlations between mean molecular weight (MMW) and temperature at 10 mbar. The left-most panel shows all retrievals which include the full suite of molecules. The middle panel shows the same but only including CO$_2$, CH$_4$ and DMS with blue and red indicating no spot and spot correction reductions respectively. The right-most panel shows all retrievals for the fiducial spectrum (Blc-EQ-S, binned before fitting light curves - empirical limb darkening - with spot correction) with blue indicating those with all molecules and red those with only CO$_2$ and CH$_4$. }
    \label{fig:MMW_T}
\end{figure*}

Interestingly, we find a general trend across our results in which higher MMWs correlate with higher retrieved temperatures (see Figure \ref{fig:MMW_T}). This trend is opposite to the known degeneracy for H$_2$-He-dominated atmospheres, where increasing abundances or temperatures both enhance the spectral features, and likely arises from the plausibility of both light- and heavy-atmosphere scenarios for this planet. In the latter, increasing MMW can have the opposite effect on spectral features by lowering the atmospheric scale height.

\subsection{Retrievals from a Literature Spectrum}
We also tested our retrieval setups on the transmission spectrum published by \cite{Madhusudhan2023} (hereafter ``Orig-Mad23'')\footnote{Downloaded from: \url{https://osf.io/36djh/}}, as well as to a rebinned version using the same wavelength bins as our extractions (``Bts-Mad23''). As in the original study, the \texttt{TauREx} retrievals find tighter constraints on the CO$_2$ abundance when using the Orig-Mad23 and Bts-Mad23 spectra, compared to any of our own extractions. The corresponding CO$_2$ detection significances are $>3\sigma$ with the simplified chemical setups, and 2–3$\sigma$ with the comprehensive setup. DMS is tentatively detected at 2.25$\sigma$ from the Orig-Mad23 only using the simplified chemical setup, while its significance drops to 1.90$\sigma$ for the Bts-Mad23 spectrum.
Notably, the \texttt{TauREx} retrievals converge to higher CH$_4$ and CO$_2$ abundances (still consistent within 1$\sigma$) than those reported in the original study, shifting toward a high-MMW scenario, similar to our spectra without spot correction.
These discrepancies are partially due to the larger priors adopted in this study, effectively allowing for purely metallic atmospheres and higher temperatures than might be physically expected. Nevertheless, retrieval tests using more constrained temperature and abundance priors, analogous to \cite{Madhusudhan2023}, still recover a bimodal solution that includes the high-MMW mode. Additional subtle differences may arise from the use of different retrieval frameworks and/or opacity cross sections, underscoring the importance of careful code comparisons.

\subsection{Retrieval Performances}
The $\chi_{0}^{2}$ values of our retrievals fall in the range 1.17--1.58, comparable to those reported in other exoplanet atmospheric studies with JWST (e.g., \citealp{Banerjee2024}), thus indicating a satisfactory fit. Importantly, differences in $\chi_{0}^{2}$ across extractions are driven mainly by the size of the spectral error bars rather than a closer match to theoretical models; swapping error bars between spectra produces corresponding changes in $\chi_{0}^{2}$ values.

The comprehensive retrieval setups systematically yield slightly higher $\chi_{0}^{2}$ values, by 0.08–0.09, compared to the simplified chemical setups. The latter also have higher or comparable Bayesian evidences, supporting their statistical preference. Even slightly better metrics are obtained with an isothermal T-P profile and/or without Mie scattering and clouds (see Table \ref{tab:comp_retr_fid2}).

\texttt{TauREx} retrievals on the Orig-Mad23 spectrum yield $\chi_{0}^{2}=$1.08–-1.09, matching the AURA fit quality reported by \cite{Madhusudhan2023}. For the Bts-Mad23 spectrum, we obtain $\chi_{0}^{2}=$ 1.56-–1.66, slightly higher than the average from our own spectral extractions. Even replacing the error bars in the Bts-Mad23 spectrum with those from our Blc-EQ-S, the resulting $\chi_{0}^{2}$ would remain higher (1.42–-1.51), potentially suggesting a better agreement between our fiducial spectrum and theoretical models.

\subsection{Retrievals with Inflated Error Bars}
As an additional test, we performed retrievals with inflated spectral error bars, scaled to enforce $\chi_{0}^{2}=1$ \citep{Benneke2013,Lueber2025}.
In this case, the overall atmospheric picture remains largely unchanged, with CH$_4$ detected at $>3\sigma$, while the significance of CO$_2$ and DMS drops below 3$\sigma$ even in the Bts retrievals (see Table \ref{tab:retr_largerr}).

\section{Implications for Atmospheric Retrievals}

\subsection{Stellar Spot and Spectral Offsets}

At zeroth order, the spot correction primarily introduces a uniform offset in the NIRISS spectrum: +15 ppm (Blc-EQ-S relative to Blc-EQ) or +24 ppm (Blc-PhQ-S/Blc-PhQ-S$'$ relative to Blc-PhQ), with a standard deviation of 9 ppm in the corresponding difference spectra (see Figure \ref{fig:transpec_corner}). For comparison, the mean spectral error bar at $R=100$ is $\sim$70 ppm. 

In Appendix \ref{app:transpec_offset}, we show that the mean offset between the NIRISS and NIRSpec spectra largely explains the different interpretations found when including spot corrections or not. In short, spectra offsets are the main driver of the variations in the retrieved CH$_4$ abundance and atmospheric MMW for this K2-18\,b data. We also find that the \texttt{TauREx} retrievals statistically prefer our Blc-EQ-S and Blc-PhQ-S spectra as-is--- i.e., without additional offsets (see Table \ref{tab:retr_offset})---suggesting that the spot correction might be performing effectively. Moreover, a comparison between multiple retrieval configurations can be seen in Figure \ref{fig:multi_posteriors}.

However, we note that offsets in transmission and eclipse spectra in multi-epoch and/or instrument observations can arise from multiple sources, including instrumental systematics and inter-visit variability of any kind, and can influence the retrieved molecular abundances and atmospheric properties. This effect can be mitigated if each dataset has resolvable spectral features and some overlap between them \citep{Edwards2024,Edwards2023}. Nevertheless, a marginally better retrieval fit does not constitute conclusive evidence that such offsets are absent.

\subsection{Understanding Trends with Retrieval Complexity}
A clear outcome of our thorough analysis of K2-18\,b is that assumed model complexity plays a critical role in shaping atmospheric inferences. In particular, at low SNR such as with this K2-18\,b data, retrieval conclusions are dependent on the number of molecules included in the fit. Our simplified retrieval setups systematically lead to higher molecular detection significances than the comprehensive ones, based on direct model comparisons with the leave-one-out approach. This effect is particularly important for establishing the presence of certain absorbers. For instance, CO$_2$ and DMS are marginally supported only in the simplified setups (at 2–3$\sigma$), with DMS further restricted to a few specific spectral extractions. Their vanishing significance in more comprehensive setups indicates that alternative explanations can account for the data. Differences in log-evidence alone may lead to fake detections if the range of plausible models is not adequately explored. This might be the case for the initial claims of H$_2$O vapour in K2-18\,b from HST/WFC3 spectra \citep{Benneke2019,Blain2021}, as well as for other controversial results in the exoplanet literature (e.g., \citealp{Evans2016,Merritt2020,Welbanks2025}).

Beyond detection significances, we noted that the number of molecules affects the retrieved abundances, with comprehensive setups yielding higher CH$_4$ and atmospheric MMW (see Sections \ref{sec:retr_fid} and \ref{sec:retr_other}). It may be surprising that molecules lacking notable spectral signatures can influence retrieval results, despite broad priors allowing them to remain minimal. A careful examination shows that most undetected molecules are not strongly ruled out by the data, permitting only loose upper bounds on their abundances. We hypothesize that these molecules contribute spectral noise and/or raise the MMW (thus reducing the atmospheric scale height), thereby pushing retrievals toward higher CH$_4$ abundances to match the observed features.

In summary, the choice of molecules in a retrieval setup can bias results in multiple ways. Including too few species may lead to spurious detections and bias the inferred parameters as the model compensates for missing inputs. Conversely, including too many species can also introduce biases, particularly when their abundances are poorly constrained by the data. Our approach of systematically comparing many retrieval configurations provides a practical compromise, capturing the key atmospheric signals while minimizing biases from both extremes.

\subsection{Comparison with Previous Studies}

\begin{figure*}[ht]
    \centering
    \includegraphics[width=\textwidth]{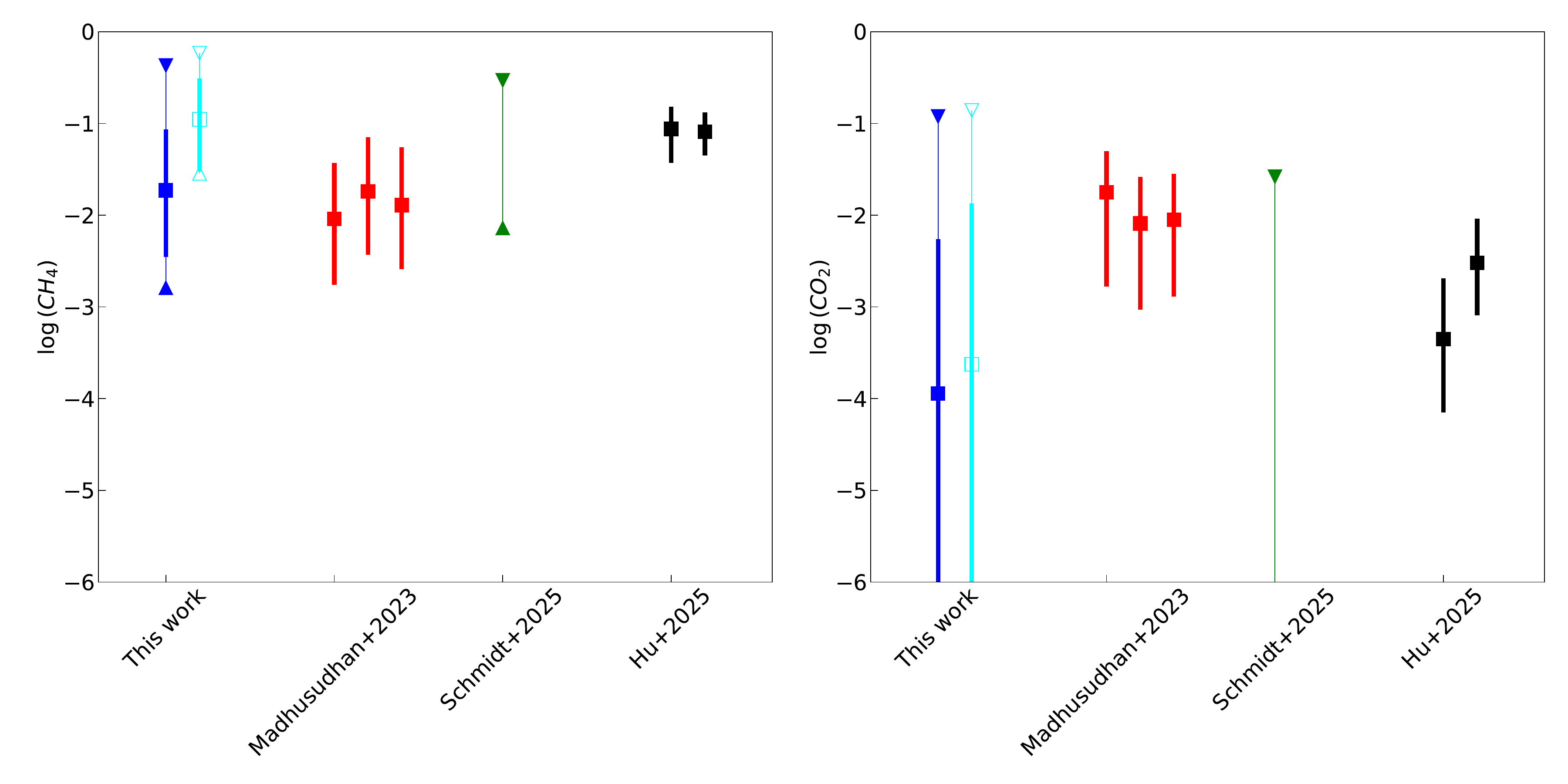}
    \caption{Atmospheric abundances of CH$_4$  and CO$_2$ obtained in this work, compared with those reported by other papers. This work: blue denotes simplified retrieval, cyan is for comprehensive retrieval results; the thicker error lines are derived from the fiducial Blc-EQ-S spectrum, the thinner lines denote the 1$\sigma$ range covered by all spectral extractions. From \cite{Madhusudhan2023}: three values from their Table 2, obtained with zero, one or two free offsets between detectors. From \cite{Schmidt2025}: total 1$\sigma$ range for CH$_4$ and 2$\sigma$ upper limit across multiple reductions, as reported in their Section 4.2.2. From \cite{Hu2025}: Results from their ExoTR retrievals (Table 5, Shifted Average) and Aura retrievals (Table 7, Hazes). Notably, all results are consistent within about 1-2$\sigma$.}
    \label{fig:retr_comp_papers}
\end{figure*}

Our reference analysis (Blc-EQ-S extraction + simplified retrieval) recovers a CH$_4$ abundance very similar to the three initial values reported by \cite{Madhusudhan2023}, while our CO$_2$ posterior distribution has a median two orders of magnitude smaller and a long tail toward negligible values, though still consistent within 1$\sigma$. Considering all our retrieval results, the conservative 1$\sigma$ range for CH$_4$ is slightly broader than that obtained by \cite{Schmidt2025}, whereas the 1$\sigma$ upper limit for CO$_2$ exceeds their 2$\sigma$ upper limit. The latest study by \cite{Hu2025}, incorporating new datasets, reports CH$_4$ and CO$_2$ abundances that fall in the upper halves of our ranges (most closely matching the CH$_4$ range from our comprehensive retrievals), while preserving a CO$_2$-to-CH$_4$ ratio similar to ours.

Regarding DMS, we find that its detection is marginal and limited to specific extraction and retrieval configurations, consistent with the results reported in \cite{Schmidt2025}.

\section{Implications for Planet Formation}

\subsection{Overview of Formation Theories}

The origin of the close-in Super-Earth / Sub-Neptune planets remains under debate. Formation theories can be divided into two broad classes: 1) Formation in the outer disk, followed by tidal migration to the inner region \citep[e.g.,][]{2012ARA&A..50..211K,2015A&A...582A.112B,2021A&A...656A..69E}; 2) Formation {\it in situ} in the inner region \citep[e.g.,][]{2012ApJ...751..158H,2013ApJ...775...53H,2013MNRAS.431.3444C,2014ApJ...780...53C}. 

The composition of the planetary ``core'' and atmosphere are likely to depend on the location in the protoplanetary disk where formation occurred. Note, here we define ``core'' as the solid/liquid component of the planet that is not part of the atmosphere. Planetary cores that are assembled, either by planetesimal or pebble accretion, in the cooler, outer disk, especially beyond the water ice line at $\sim 170\:$K, are expected to have bulk compositions that are rich in volatile species. This would lead to the core having significant amounts of water such that a global planetary ocean is a significant fraction of the mass budget. 

On the other hand, if the planetary core is assembled from small solids, i.e., pebbles or planetesimals, in the warm, inner regions interior to the water ice line, then its bulk composition is expected to be mainly composed of metals and metal-rich silicates \citep[e.g.,][]{2022MNRAS.517.2285C}. Refractory interstellar dust also contains a carbonaceous component, often referred to as ``soot'', which is significant in terms of the global carbon budget, although likely to be sub-dominant in terms of mass compared to the metals/silicates \citep[e.g.,][]{2025arXiv250816781L}. This soot is expected to sublimate at temperatures near $\sim500\:$K, with this location in the protoplanetary disk referred to as the ``soot line''. Thus, planetary cores are expected to become relatively carbon poor if formed interior to the soot line, although the overall impact on bulk density may be relatively modest.

If core masses become sufficiently massive ($\gtrsim 1\:M_\oplus$) while the planet is still embedded in a gas-rich disk, then accretion of a ``significant'' H/He-rich ``primordial'' atmosphere becomes possible. The ability of the planetary core to accrete H/He gas depends on local environmental conditions and the ability of the gas to cool. For example, even at temperatures of $\sim1000\:$K, planetary cores of $\sim 1\:M_\oplus$ may accrete a few percent by mass of H/He (Thomopoulos et al., in prep.). Such an atmosphere, while a very minor component of the total mass, causes the size of the planet, i.e., defined by its optically thick photosphere, to be much larger than the core surface. Under cooler conditions and for more massive cores, the mass fraction of the primordial atmosphere is expected to increase. Finally, we note that for core masses greater than a certain critical value, $\sim 10\:M_\oplus$, runaway gas accretion leading to gas giant formation is expected \citep[e.g.,][]{1996Icar..124...62P,2014ApJ...786...21P}.

If dominated by H/He, a primordial atmosphere will have a low MMW, i.e., 2.33 amu in the limit of pure $\rm H_2$ and He with $n_{\rm He}=0.2n_{\rm H2}$. The primordial atmosphere's abundance of heavier elements will also reflect the composition of gas in the local zone of the protoplanetary disk. For example, in astrochemical models of protoplanetary disks that including radially drifting pebbles, enhanced gas phase water is expected interior to the water ice line \citep[e.g.,][]{2022MNRAS.517.2285C}. Indeed, elevated abundances of gas phase water have been reported in the inner regions of some disks \citep[][]{2023ApJ...957L..22B}, which is interpreted as indirect evidence for pebble drift. However, in addition to gas accretion, the composition of the primordial atmosphere could be affected by continued pebble and/or planetesimal accretion, which would lead to enhanced abundances of refractory elements.

Once a primordial atmosphere has formed, its composition may be further altered by mass loss due to photoevaporation, especially if this tends to preferentially remove lighter species, i.e., H and He. Indeed, \citet{2021MNRAS.503.1526R} have modelled the Super-Earth / Sub-Neptune population as having common properties of $\sim$few Earth-mass rocky/iron-rich cores with initial primordial atmospheres of a few percent by mass, with the Super-Earths, i.e., below the ``radius valley'', being revealed by photoevaporation of their atmospheres. Interaction with the planetary core could also alter the composition of a primordial atmosphere. For example, atmospheric material may dissolve in a global ocean. And/or volatile materials may be outgassed from the core \citep[e.g.,][]{2025arXiv250402499H}. In the limiting case where this outgassed component dominates, then this ``secondary'' atmosphere would no longer reflect the original gaseous composition of the local protoplanetary disk.

Given the above possibilities, it is in general difficult to make definitive predictions of atmospheric composition that are unique indicators of a given formation location in the protoplanetary disk. On the other hand, for a given formation scenario, where both the location and mode of accretion are specified, we may attempt to make predictions for the atmospheres of its resulting planets. In the next sub-section we outline such predictions for the {\it in situ} model Inside-Out Planet Formation \citep{2014ApJ...780...53C}.

\begin{figure*}[ht]
    \centering
    \includegraphics[width=\textwidth]{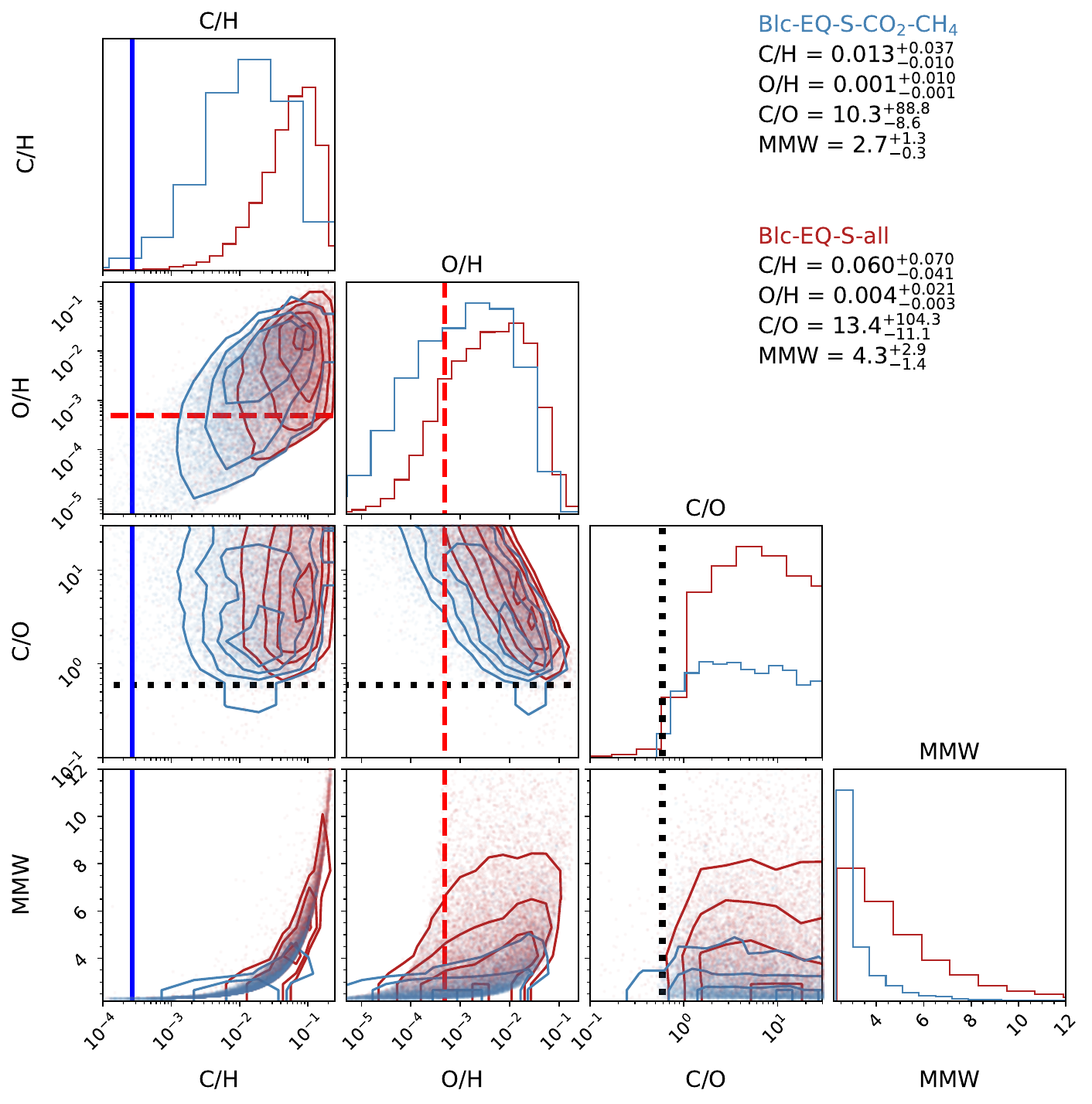}
    \caption{Derived posterior distributions of carbon and oxygen abundances with respect to hydrogen, elemental C/O ratio and mean molecular weight (MMW) for the fiducial reduction. Red shows the comprehensive retrieval and blue the simplified retrieval. The plots are zoomed in to highlight regions of interest such as C/O $\sim 0.1-30$. The solid blue line is the solar C/H, the dashed red line the solar O/H, and the black dotted line the solar C/O. On the top right, the median and 1$\sigma$ uncertainties for both reductions are shown.}
    \label{fig:c_to_o}
\end{figure*}

\subsection{K2-18\,b and c in the context of Inside-Out Planet Formation}

Inside-Out Planet Formation (IOPF) \citep{2014ApJ...780...53C} is a model for the {\it in situ} formation of close-in Super-Earth/Sub-Neptune multi-planet systems. Planets form from pebble-rich rings that are trapped at the pressure maximum associated with the dead zone inner boundary (DZIB) with an interior magneto-rotational-instability (MRI) active zone, with this location first set by thermal ionization of alkali metals at about 1,200~K. The location of the DZIB, i.e., where the disk midplane temperature reaches about 1,200~K, is estimated to be:
\begin{equation}
    r_{\rm DZIB}=0.0653\phi_{\rm DZIB,0.5}\gamma_{1.4}^{-2/9}\kappa_{10}^{2/9}\alpha_{-3}^{-2/9}m_{*,1}^{1/3}\left(f_r\dot{m}_{-9}\right)^{4/9}\ {\rm au},
    \label{eq:r_dzib}
\end{equation}
where $\phi_{\rm DZIB,0.5}\equiv\phi_{\rm DZIB}/0.5$ is a dimensionless parameter of order unity accounting for potential differences from a pure viscous disk model (especially expected to be somewhat less than unity due to energy extraction by a disk wind), $\gamma\equiv1.4\gamma_{1.4}$ with fiducial normalization to a
value of 1.4 for $\rm H_2$ with rotational modes excited, $\kappa_{10}
\equiv \kappa / 10\ {\rm cm^2\:g^{-1}}$ is the disk mean opacity with fiducial normalization appropriate for inner disk
conditions, $\alpha_{-3}\equiv\alpha/10^{-3}$ is the Shakura-Sunyaev viscosity parameter with normalization guided by numerical simulations of dead zone regions of protoplanetary disks, $m_{*,1}\equiv
m_*/M_\odot$ is the stellar mass with fiducial normalization of
$1\:M_\odot$, $f_r \equiv1-\sqrt{r_*/r_{\rm DZIB}}$, $r_*$ is the stellar radius,
and $\dot{m}_{-9}\equiv\dot{m}_*/10^{-9}\:M_\odot\:{\rm yr}^{-1}$ is the disk accretion rate. 

The innermost planet in the K2-18 system is K2-18\,c. It is a non-transiting Super-Earth detected via radial velocity measurements \citep{Cloutier2017,Cloutier2019}. A later analysis by \cite{Radica2022b} using the line-by-line framework confirmed the presence of the planet and revised the mass and period. Nevertheless, \cite{Radica2022b} do not provide values for semimajor axis so we opt to use the values from \cite{Cloutier2019} for semimajor axis. These are a minimum mass of $6.99^{+0.96}_{-0.99}\,\mathrm{M_{\oplus}}$\footnote{The value obtained by \cite{Cloutier2019} is $5.62\pm0.84\,\mathrm{M_{\oplus}}$}, semi-major axis of $0.0670\:\pm0.0002\:$au and equilibrium temperature of $409\pm8$\,K assuming a bond albedo of 0.3.

If the current location of K2-18\,c, i.e., $r=0.067\:$au, is the location of the DZIB, then, given the host star mass of $m_*=0.495\:M_\odot$, this implies $\dot{m}_{-9} = 
%0.567\phi_{\rm DZIB,0.5}^{-9/4}\alpha_{-4}^{-1/2} = 
1.80\phi_{\rm DZIB,0.5}^{-9/4}\alpha_{-3}^{1/2}$, where $f_r$ and other factors have been assumed to be unity. We note that this accretion rate, i.e., $\sim10^{-9}\:M_\odot\:{\rm yr}^{-1}$, is typical of those observed in transition disk systems \citep{2014A&A...568A..18M}, i.e., where dust appears to be depleted from the innermost regions. This derived condition on accretion rate and viscosity in the vicinity of the DZIB also places a constraint on the size of the pebbles that are trapped at this location, i.e., pebble radius $a_p>0.15\:$cm \citep{2025arXiv250810755H}.

The planet mass in the IOPF theory is assumed to be the gap opening mass. From IOPF Paper IV \citep{2018ApJ...857...20H}, this is given by
\begin{equation}
M_{p,1}=11.1 \phi_{\rm G,D,1.44} \phi_{\rm DZIB,0.5}^{-9/8}\gamma_{1.4}^{5/4}\alpha_{-3}^{1/2}m_{*,1}^{-1/4}r_{\rm 0.1au}^{5/4}\:M_\oplus,
\end{equation}
where $\phi_{G,D,1.44}\equiv\phi_{G,D}/1.44$ is a dimensionless factor with fiducial value of unity that has been calibrated from numerical simulations. Again, evaluating for K2-18\,c parameters, yields 
$M_{p,1}=
%2.53\phi_{\rm DZIB,0.5}^{-9/8}\alpha_{-4}^{1/2}\:M_\oplus = 
8.00\phi_{\rm DZIB,0.5}^{-9/8}\alpha_{-3}^{1/2}\:M_\oplus$. 
%$M_{p,1}=7.45\phi_{\rm DZIB,0.5}^{-9/8}\alpha_{-4}^{1/2}\:M_\oplus$.
Comparing to the observed mass of
K2-18\,c, i.e., $M_{p,1,{\rm obs}}\simeq 6.99^{0.96}_{0.99}\:\mathrm{M_{\oplus}}$ 
%(revised value from 
\citep{Radica2022b} (note, we assume sin $i \simeq 1$), 
%K2-18\,b, i.e., $M_{p,1,{\rm obs}}=8.63\pm1.35\:M_\oplus$, 
we see that the fiducial IOPF predicted mass is consistent with K2-18\,c's properties if $\alpha\simeq10^{-3}$.
%$\alpha_{-4}=8.81$.

The predicted mass of K2-18\,b can be estimated from that of the innermost planet given the scaling of gap opening mass:
\begin{equation}
    M_{\rm G,D} = 6.86 \phi_{\rm G,D,1.44} \gamma_{1.4} \kappa_{10}^{1/4}\alpha_{-3}^{1/4}m_{*,1}^{1/8}(f_r\dot{m}_{-9})^{1/2} r_{\rm 0.1au}^{1/8} M_\oplus,
\end{equation}
i.e., the gap opening mass scales as $r^{1/8}$. Thus, given the properties of K2-18~c and the location of K2-18~b at 0.159~au, the latter's mass is predicted to be $(6.99\pm0.98) \times(0.159/0.067)^{1/8}M_\oplus=7.8\pm1.1\:M_\oplus$. This is consistent with the observed value of $8.63\pm1.35\:M_\oplus$.

Next we consider the implications of an IOPF formation scenario for the composition of the core and atmosphere of K2-18\,b. In IOPF, the solid core of the planet would be built up by pebbles that have survived to the local disk temperature at K2-18~b's location. For a viscously heated disk, the midplane temperature is predicted to scale as $T\propto \gamma_{1.4}^{-1/5}\kappa_{10}^{1/5}\alpha_{-3}^{-1/5}\dot{m}^{2/5}r^{-9/10}$. The location of K2-18~c sets the place in the disk where the midplane temperature was $\sim1,200\:$K. Thus, the local disk temperature of K2-18~b during formation is expected to be $\sim 550\:$K. This temperature, which we note is much warmer than the current equilibrium temperature, is expected to be just interior to the ``soot line'', which has fiducial expected temperature of $\sim500\:$K \citep[e.g.,][]{2021SciA....7.3632L}. Thus the core of the planet is expected to be volatile poor, i.e., mainly metal/rock-rich composition, with low water and carbon content. 
%, i.e., would be silicate and iron rich and would be relatively volatile poor. 
However, because the planet is embedded in a gas-rich disk, it has the chance to accrete a primordial atmosphere, with typical mass fractions of $\sim 1$ to 10\%. These conditions for the solid cores and initial primordial atmospheres have been inferred for the general Super-Earth/Sub-Neptune population by the photo-evaporation modelling study of \citet{2021MNRAS.503.1526R}. In this context, for K2-18\,b if the core has a mass of $\sim8\:M_\oplus$ average density of $\sim 6\:{\rm g\:cm}^{-3}$, then its radius would be $1.94\:R_\oplus$. Given the observed radius of K2-18~b of $2.61\pm0.09\:R_\oplus$, the atmosphere needs to have a thickness of about $35\%$ of the core radius. Such a primordial atmosphere, retained since the time of formation of the planet, would be in the regime of being a low mean molecular weight atmosphere.

Detailed predictions for the atmospheric composition of the innermost planets forming via IOPF have not yet been made. IOPF Paper VII \citep{2022MNRAS.517.2285C} presented chemodynamical models of protoplanetary disks, coupling a gas-grain astrochemical network with pebbles undergoing radial drift. They found enhanced abundances of gas-phase $\rm H_2O$ were delivered to regions inside the water ice line at $T\sim 170\:$K. This could boost the gas-phase O abundance by factors of several, leading to gas-phase $\rm C/O$ ratios of $\sim 0.1$. In IOPF the primordial atmospheric composition is expected to reflect that of the gas-phase abundances in the disk, since pebble accretion would be truncated by the DZIB that has retreated further out in the disk. In this case, planets just interior to the water-ice line would have low C/O ratios. In the case of the K2-18 system, scaling from the innermost planet that is assumed to be at 1,200~K, the location of the 170~K water ice line is $r_{\rm 170K}=0.587\:$au.

However, as mentioned above, at a temperature of about 500~K, the carbonaceous ``soot'' component of dust is expected to sublimate. The location of this soot line in the K2-18 system is at $r_{\rm 500K}=0.177\:$au, so K2-18\,b is expected to be just interior to this soot line. From stoichiometry, the abundance of ``soot'', assuming it dominates the refractory C content of dust, is expected to be about $[{\rm C_{soot}/H}]\sim 1.6\times 10^{-4}$, i.e., about $10\times$ greater than that of water ice \citep{2022MNRAS.517.2285C}, so this would boost the C/O ratio to values $\gtrsim 1$. Thus, the prediction of the IOPF model for the atmospheric composition of K2-18~b is a primordial, i.e., low MMW, atmosphere with an elevated C/O ratio.

In Figure~\ref{fig:c_to_o} we highlight several metrics derived from our atmospheric retrievals of K2-18~b: C/H; O/H; C/O; and MMW. We see that low MMW solutions are preferred, favouring the scenario of a primordial H/He dominated atmosphere. An abundance of O that is elevated above solar, typically by a factor of 10, is the best solution, which would be consistent with gas that has been enriched in water inside the water ice line \citep{2022MNRAS.517.2285C}. However, the retrieved C abundance is even more enhanced above solar, i.e., by factors of $\sim 100$, which is consistent with the expected composition of protoplanetary disk gas interior to the soot line.

The above conclusions are consistent with a number of recent studies of K2-18~b. For example, \citet{2024A&A...686A.131L} presented a detailed model of the structure of K2-18~b's atmosphere, concluding it was unlikely to be able to host a liquid water ocean below a H/He dominated atmosphere. \citet{2025arXiv250910947L} have recently presented results of atmospheric retrievals of K2-18~b based on JWST NIRISS/NIRSpec and MIRI LRS data, and allowing for the presence of hazes. Their preferred solutions are low MMW ($\sim2.4\:$amu) atmospheres with the presence of hydrocarbon hazes, whose formation would tend to indicate an elevated C/O ratio.

\section{Conclusions}

In this work we re-reduced and analysed the original NIRISS and NIRSpec JWST observations of the sub-Neptune K2-18\,b. We have made use of publicly available pipelines in combination with custom-made pipelines producing a total of 12 transmission spectra, exploring the effects of different treatments of limb darkening, spectral binning, error inflation, instrumental offsets, and spot correction. Notably, we introduced a new semi-empirical method for correcting stellar spot crossing events. We also performed retrievals on the transmission spectrum from \cite{Madhusudhan2023} and our binned version of it. 

We recommend extracting a robust transmission spectrum by using binned light curves (R$\sim$100) to mitigate correlated-parameter biases in the low-S/N regime, adopting empirical limb-darkening coefficients, and, where applicable, applying our novel semi-empirical spot correction method to minimize model-imposed biases. The validity of this approach is further supported by our statistical tests and is consistent with lessons learned in previous studies by the community \citep{Carter2024,Davey2025}. Binning is particularly effective to enhance the definition of broad features, such as those of CH$_4$ and CO$_2$ present in this atmosphere. Nevertheless some narrow or small-amplitude features may be lost when binning in wavelength, depending on the atmospheric composition and S/N of the data, therefore the optimal resolution or choice of bins may be case-dependent.

We also propose a suite of retrieval configurations to test atmospheric model dependencies, including different clouds, hazes, and temperature–pressure parametrizations, both comprehensive and minimal sets of chemical species, and criteria for selecting a simplified retrieval setup. We detect $\mathrm{CH_4}$ with a significance of above 3--4$\sigma$ in all configurations. The detection of $\mathrm{CO_2}$ is configuration dependent and, in most cases, around 2$\sigma$. The absolute abundances of both species are broadly constrained and slightly dependent on the configuration, encompassing both scenarios with a primordial H$_2$-He atmosphere or a heavier secondary atmosphere. However our fiducial solution favours the first scenario.

Finally, by computing the C/H, O/H, and C/O ratios, we find that K2-18 b exhibits super-solar abundances of both carbon and oxygen, with a much stronger enhancement in carbon, resulting in a significantly super-solar C/O ratio. These atmospheric properties are in line with expectations of the Inside-Out Planet Formation (IOPF) model, if K2-18~b formed interior to the ``soot'' line. We have also shown that the physical properties (masses and orbital locations) of the innermost planet, K2-18~c, and K2-18~b are consistent with the IOPF scenario. This is the first test of IOPF via examination of both physical planetary properties and atmospheric chemical composition. Clearly larger samples of planets with measured atmospheric compositions are needed, ideally in multi-planet systems that span a range of disk conditions that straddle the soot and water ice lines.

\begin{acknowledgements}
We thank Arturo Cevallos Soto, Kevin Heng, Greg Houlihan, Xiao Hu and Antonino Petralia for helpful discussions. G.F-R. and E.Po. acknowledge support from Chalmers Astrophysics and Space Sciences Summer (CASSUM) fellowships. G.M. acknowledges financial support from the Severo Ochoa grant CEX2021-001131-S and from the Ramón y Cajal grant RYC2022-037854-I funded by MCIN/AEI/1144 10.13039/501100011033 and FSE+. J.C.T. acknowledges support from the Chalmers Initiative on Cosmic Origins (CICO) and the Virginia Institute for Theoretical Astrophysics (VITA), supported by the College and Graduate School of Arts and Sciences at the University of Virginia. This work is based on observations made with the NASA/ESA/CSA James Webb Space Telescope obtained from the Mikulski Archive for Space Telescopes (MAST) at the Space Telescope Science Institute (STScI). STScI is operated by the Association of Universities for Research in Astronomy, Inc., under NASA contract NAS 5–03127.

\end{acknowledgements}

\bibliographystyle{aa}
\bibliography{ref}

\begin{appendix}

\section{Comparison of Transmission Spectra}
\label{app:transpec_comp}

\begin{figure}[ht]
    \centering
    \includegraphics[width=\textwidth]{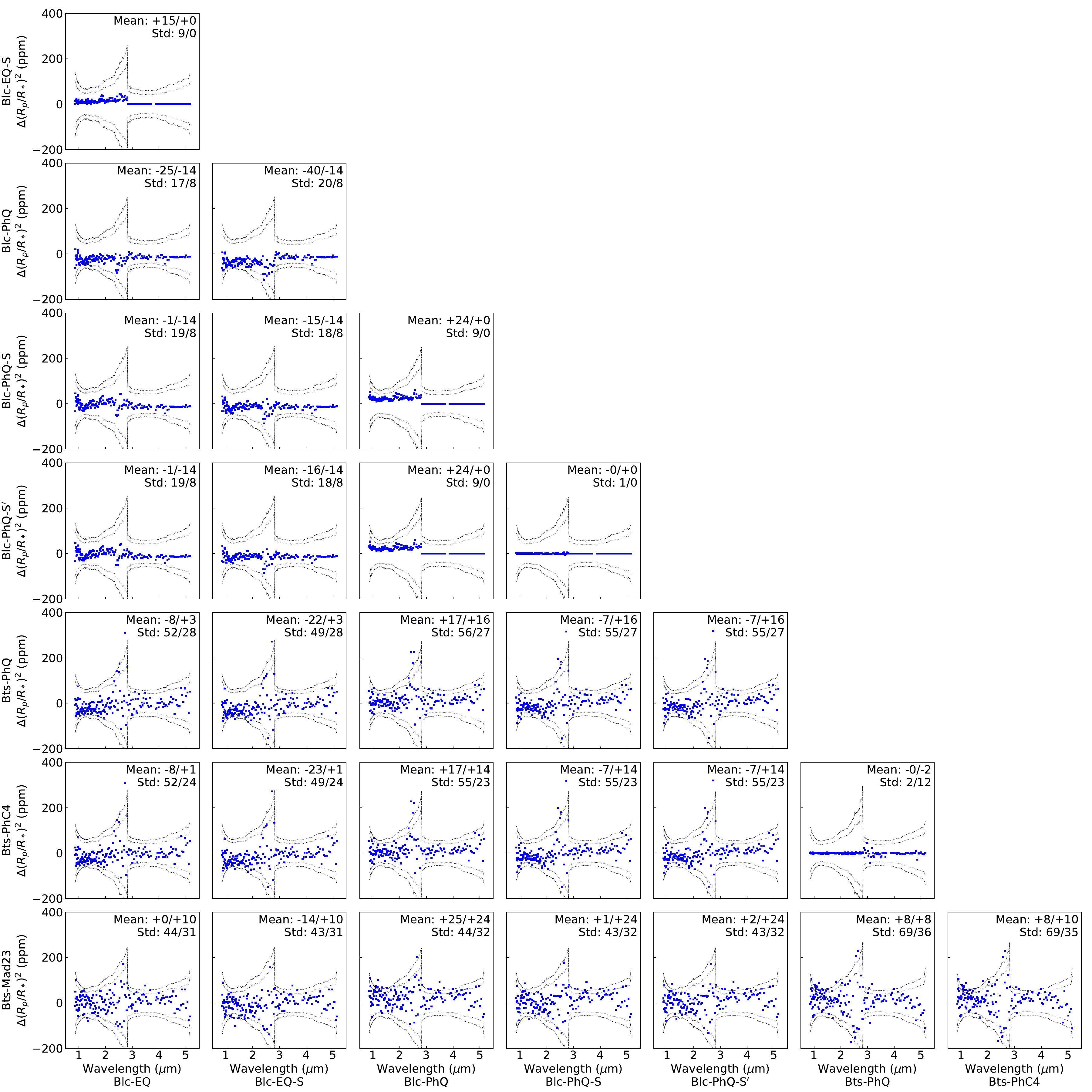}        \captionsetup{width=\textwidth}
    \caption{Cornerplot of difference between transmission spectra extracted with various methods (see Section \ref{sec:spec_lc_fits}).}
    \label{fig:transpec_corner}
\end{figure}

Figure \ref{fig:transpec_corner} presents a corner plot of all pairwise differences between the transmission spectra from various methods in this work (see Sections \ref{sec:spec_lc_fits} and \ref{sec:fid_spec_lc}) and the spectrum from \cite{Madhusudhan2023}, binned to a common wavelength grid. Transmission spectra from all our Blc setups agree well within 1$\sigma$, though stellar spot and limb-darkening treatments may lead to subtle systematic differences.
For instance, the spot correction shifts the average NIRISS transit depth upward by 15 ppm or 24 ppm compared to the uncorrected cases when using empirical or \texttt{PHOENIX} quadratic limb-darkening coefficients, respectively, with maximum differences reaching 45 ppm or 62 ppm. Accounting for the wavelength-dependent spot signal shape due to limb-darkening alters transit depths by at most 5 ppm.
The difference spectra between extractions using empirical and \texttt{PHOENIX} limb-darkening  are more randomly scattered, though both instruments show a consistent mean offset of $\sim$+15 ppm when using empirical rather than \texttt{PHOENIX} coefficients.

Differences between Blc and Bts transmission spectra exhibit greater scatter, with some data points deviating by more than 1$\sigma$. The largest discrepancies occur near the detector edges, consistent with previous technical reports (e.g., \citealp{Carter2024}). In particular, the 2.5–4\,$\mu$m region, corresponding to the NIRISS–NIRSpec transition, is critical for interpreting atmospheric composition. This range includes CO$_2$ and DMS features, which are likely driving the different retrieval outcomes (see Section \ref{sec:retr_other}).

We also compare our transmission spectra with the first reduction published by \cite{Madhusudhan2023}, rebinned to wavelength bins adopted in this study (hereafter ``Bts-Mad23''). The Bts-Mad23 spectrum is morphologically closer to our Blc spectra, with standard deviations of the differences of $\sim$43 and 31 ppm for the NIRISS and NIRSpec segments, respectively. In comparison, the analogous values with our Bts spectra are 69 and 36 ppm. Additionally, offsets between our spectral segments and Bts-Mad23 range from –14 to +25 ppm; these offsets are similar across instruments when no spot correction is applied, (Blc-EQ, Blc-PhQ, Bts-PhQ and Bts-PhC4) but differ more when our spot correction is included. %(Blc-EQ-S,  Blc-PhQ-S, Blc-PhQ-S^\prime$).
\clearpage
\onecolumn

\clearpage
\twocolumn

\clearpage
\onecolumn

\section{Retrievals with Alternative T–P Profiles, Clouds and Haze Treatments}
\label{app:retr_tp_cloud_haze}

Table \ref{tab:comp_retr_fid2} summarises the results from different retrievals on the fiducial spectrum using various temperature profiles with and without clouds/hazes.

\begin{table*}[!ht]
\caption{Selected parameters from retrievals on the fiducial transmission spectrum (Blc-EQ-S).}            % title of Table
\label{tab:comp_retr_fid2}      % is used to refer this table in the text
\centering                                      % used for centering table
\begin{tabular}{ccccccc}          % centered columns (4 columns)
\hline\hline
Setup & $\chi_0^2$ & $\log{\mathrm{ev}}$ & $\log{(\mathrm{CO}_2)}$ & $\log{(\mathrm{CH}_4)}$ & $T_{\mathrm{10 \, mbar}}$ (K) & $\log{(P_\mathrm{cloud}/\mathrm{bar})}$ \\
\hline
\textbf{Gui-Cl-Mie} & \textbf{1.17} & $\textbf{1448.35}$ & $-\textbf{3.94}_{-4.20}^{+1.68}$ \textbf{(2.00$\sigma$)} & $-\textbf{1.73}_{-0.73}^{+0.66}$ \textbf{(4.62$\sigma$)} & $\textbf{138}_{-39}^{+48}$ & $\textbf{-1.14}_{-0.59}^{+1.35}$ \textbf{(N)} \\
Gui-Cl-0 & 1.14 & 1449.34 & $-4.30_{-4.90}^{+1.84}$ (1.97$\sigma$) & $-1.86_{-0.82}^{+0.80}$ (4.63$\sigma$) & $131_{-41}^{+50}$ & $-1.12_{-0.60}^{+1.42}$ (1.96$\sigma$) \\
Gui-0-Mie & 1.16 & 1449.02 & $-3.92_{-4.13}^{+1.91}$ (2.11$\sigma$) & $-1.89_{-0.97}^{+0.79}$ (4.52$\sigma$) & $108_{-32}^{+42}$ & -- \\
Gui-0-0 & 1.14 & 1448.43 & $-3.56_{-3.92}^{+1.84}$ (2.32$\sigma$) & $-1.99_{-1.07}^{+1.01}$ (5.53$\sigma$) & $103_{-30}^{+52}$ & -- \\
Iso-Cl-Mie & 1.15 & 1449.63 & $-4.18_{-4.97}^{+1.75}$ (2.05$\sigma$) & $-1.84_{-0.78}^{+0.68}$ (4.59$\sigma$) & $141_{-31}^{+39}$ & $-1.04_{-0.69}^{+1.68}$ (N) \\
Iso-Cl-0 & 1.12 & 1449.39 & $-3.81_{-5.53}^{+2.04}$ (2.19$\sigma$) & $-1.48_{-0.96}^{+0.83}$ (4.62$\sigma$) & $162_{-47}^{+95}$ & $-1.11_{-0.82}^{+1.79}$ (N) \\
Iso-0-Mie & 1.14 & 1449.41 & $-4.10_{-4.10}^{+1.87}$ (2.22$\sigma$) & $-2.05_{-0.90}^{+0.89}$ (4.77$\sigma$) & $122_{-25}^{+43}$ & -- \\
Iso-0-0 & 1.17 & 1448.92 & $-2.84_{-4.25}^{+1.46}$ (2.44$\sigma$) & $-1.39_{-1.33}^{+0.93}$ (5.30$\sigma$) & $142_{-41}^{+140}$ & -- \\
\hline
\end{tabular}
\tablefoot{Varying between Guillot or Isothermal T-P profile, inclusion of a thick cloud layer, and/or Mie scattering.}
\end{table*}

\section{Retrieval Sensitivity to Instrumental Offsets}
\label{app:transpec_offset}

Table \ref{tab:retr_offset} summarises the results from different retrievals on spot and spot-less reduction configurations including uniform offsets.

\begin{table*}[!ht]
\caption{Selected parameters from retrievals on additional spectral extractions, analogous to Table \ref{tab:retr_all}.}            % title of Table
\label{tab:retr_offset}      % is used to refer this table in the text
\centering                                      % used for centering table
\begin{tabular}{cccccccc}          % centered columns (4 columns)
\hline\hline
Spectrum & Molecules & $\chi_0^2$ & $\log{\mathrm{ev}}$ & $\log{(\mathrm{CO}_2)}$ & $\log{(\mathrm{CH}_4)}$ & $T_{\mathrm{10 \, mbar}}$ (K) & $MMW$ \\
\hline
\multirow{2}{*}{Blc-PhQ-S} & CH$_4$ & 1.31 & 1450.59 & -- & $-1.95_{-0.75}^{+0.67}$ (4.56$\sigma$) & $121_{-35}^{+44}$ & $2.46_{-0.13}^{+0.56}$ \\
 & All & 1.39 & 1449.88 & $-4.76_{-5.30}^{+2.56}$ (N) & $-0.95_{-0.60}^{+0.45}$ (4.06$\sigma$) & $146_{-48}^{+89}$ & $4.26_{-1.44}^{+3.05}$ \\
\hline
\multirow{2}{*}{\shortstack{Blc-PhQ-S\\-24 / 0}} & CO$_2$+CH$_4$ & 1.36 & 1444.74 & $-1.11_{-3.89}^{+0.33}$ (2.42$\sigma$) & $-0.91_{-1.82}^{+0.55}$ (5.02$\sigma$) & $243_{-98}^{+137}$ & $9.19_{-6.85}^{+4.06}$ \\
 & All & 1.45 & 1445.70 & $-1.88_{-4.95}^{+0.85}$ (1.82$\sigma$) & $-0.47_{-0.44}^{+0.27}$ (3.48$\sigma$) & $359_{-162}^{+180}$ & $9.99_{-4.09}^{+3.69}$ \\
\hline
\multirow{2}{*}{Blc-PhQ} & CO$_2$+CH$_4$ & 1.36 & 1446.61 & $-1.74_{-4.56}^{+0.81}$ (2.30$\sigma$) & $-1.00_{-1.79}^{+0.64}$ (4.96$\sigma$) & $213_{-79}^{+155}$ & $6.73_{-4.40}^{+5.15}$ \\
 & All & 1.45 & 1447.18 & $-2.17_{-5.46}^{+1.07}$ (N) & $-0.53_{-0.51}^{+0.30}$ (3.59$\sigma$) & $283_{-132}^{+175}$ & $8.83_{-4.27}^{+3.92}$ \\
\hline \hline
\multirow{2}{*}{Blc-EQ-S} & CO$_2$+CH$_4$ & 1.17 & 1449.47 & $-3.94_{-4.20}^{+1.68}$ (2.00$\sigma$) & $-1.73_{-0.73}^{+0.66}$ (4.62$\sigma$) & $138_{-39}^{+48}$ & $2.64_{-0.28}^{+1.16}$ \\
 & All & 1.25 & 1448.35 & $-3.63_{-5.13}^{+1.75}$ (N) & $-0.96_{-0.57}^{+0.45}$ (4.16$\sigma$) & $161_{-54}^{+90}$ & 4.32$_{-1.45}^{+3.02}$ \\
\hline
\multirow{2}{*}{\shortstack{Blc-EQ-S\\-26 / 0}} & CO$_2$+CH$_4$ & 1.23 & 1443.64 & $-1.36_{-2.94}^{+0.47}$ (2.62$\sigma$) & $-1.17_{-1.65}^{+0.68}$ (4.90$\sigma$) & $211_{-74}^{+120}$ & $6.76_{-4.42}^{+4.49}$ \\
 & All & 1.31 & 1444.50 & $-1.71_{-4.24}^{+0.66}$ (1.99$\sigma$) & $-0.51_{-0.45}^{+0.29}$ (3.65$\sigma$) & $331_{-141}^{+172}$ & $9.38_{-3.62}^{+3.64}$ \\
\hline \hline
\multirow{2}{*}{Bts-PhQ} & CO$_2$+CH$_4$+DMS & 1.50 & 1443.52 & $-5.54_{-6.07}^{+4.45}$ (3.89$\sigma$) & $-1.57_{-0.89}^{+0.67}$ (5.94$\sigma$) & $134_{-37}^{+80}$ & $2.77_{-0.41}^{+5.31}$ \\
 & All & 1.58 & 1444.07 & $-1.18_{-1.60}^{+0.32}$ (2.33$\sigma$) & $-0.64_{-0.44}^{+0.33}$ (3.83$\sigma$) & $328_{-142}^{+184}$ & $10.14_{-3.39}^{+3.21}$ \\
\hline
\multirow{2}{*}{\shortstack{Bts-PhQ\\+26 / 0}} & CO$_2$+CH$_4$ & 1.44 & 1449.16 & $-4.20_{-5.12}^{+1.81}$ (1.98$\sigma$) & $-2.05_{-0.65}^{+0.62}$ (5.33$\sigma$) & $132_{-34}^{+41}$ & $2.48_{-0.14}^{+0.55}$ \\
 & All & 1.53 & 1448.15 & $-4.26_{-5.51}^{+2.26}$ (N) & $-1.19_{-0.55}^{+0.49}$ (4.46$\sigma$) & $143_{-46}^{+93}$ & $3.34_{-0.89}^{+3.17}$ \\
\hline
\end{tabular}
\tablefoot{Testing the impact of offsets between the NIRISS SOSS and NIRSpec G395H spectral intervals. We report difference in average transit depth between NIRISS and NIRSpec, from top to bottom: -4, -28, -28 ppm || -2, -28 ppm || -29, -3 ppm. Spectra with similar differences lead to more consistent retrieval results, with manual offsets compensating for the effects of spot correction and treatment of limb-darkening. This behaviour holds, only in part, when comparing Bts and Blc spectra with offsets.}
\end{table*}

\clearpage
\twocolumn

\clearpage
\onecolumn

\clearpage
\twocolumn

\clearpage
\onecolumn
\section{Retrieval Tests with Inflated Error Bars}
\begin{figure}[ht]
    \centering
    \includegraphics[width=\textwidth]{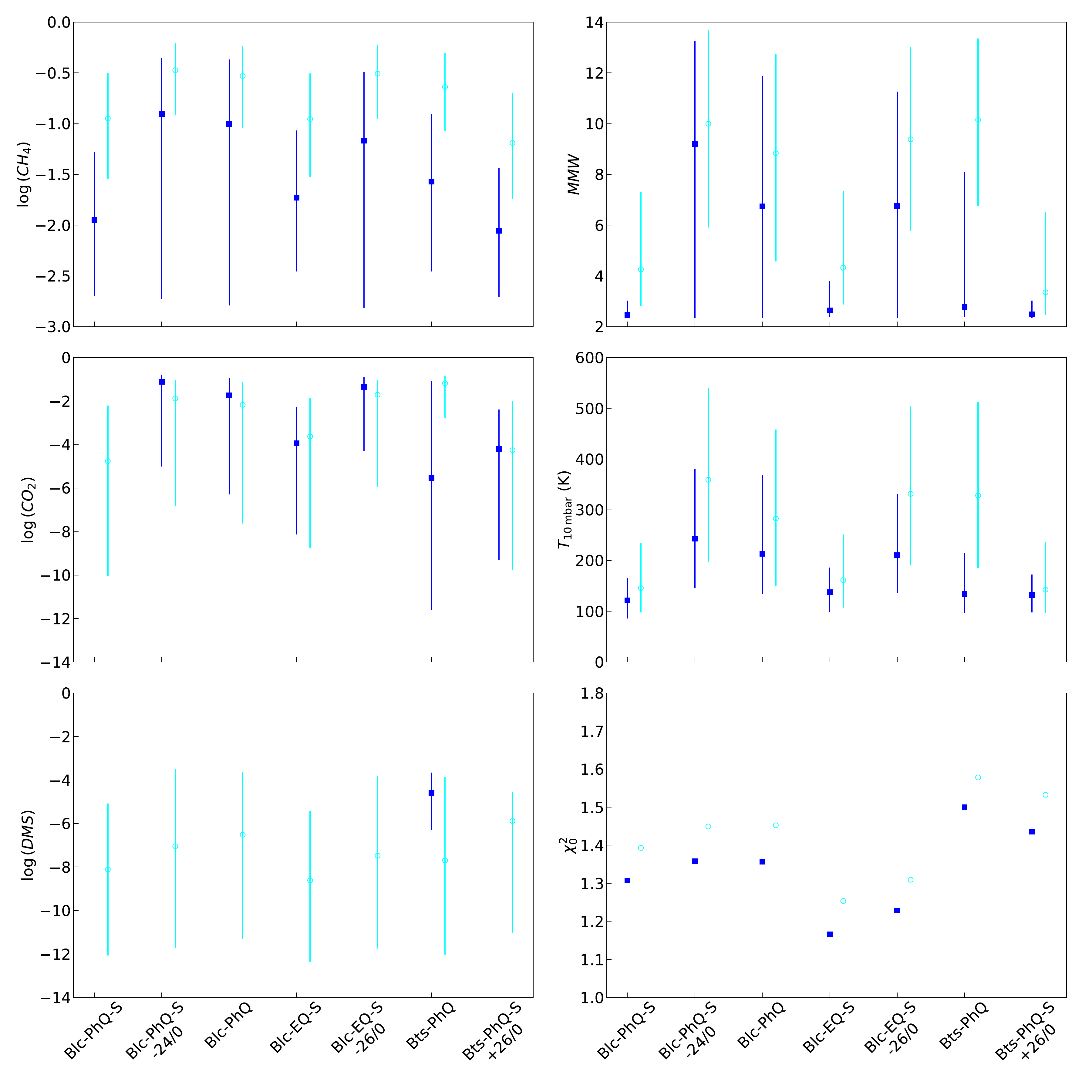}
    \caption{Analogous to Figure \ref{fig:main_retr_comp}, but comparing spectra with and without offsets (see also Table \ref{tab:retr_offset}).}
    \label{fig:offset_retr_comp}
\end{figure}

Figure \ref{fig:offset_retr_comp} shows the comparison between all different reductions with simplified and comprehensive retrievals for a number of parameters. Table \ref{tab:retr_largerr} shows the effect of inflated error bars on selected parameters and reductions.

\label{app:transpec_largerr}

\clearpage
\twocolumn

\clearpage
\onecolumn

\begin{table*}[!ht]
\caption{Selected parameters from retrievals on additional spectral extractions, analogous to Table \ref{tab:retr_all}.}            % title of Table
\label{tab:retr_largerr}      % is used to refer this table in the text
\centering                                      % used for centering table
\begin{tabular}{cccccccc}          % centered columns (4 columns)
\hline\hline
Spectrum & Molecules & $\chi_0^2$ & $\log{\mathrm{ev}}$ & $\log{(\mathrm{CO}_2)}$ & $\log{(\mathrm{CH}_4)}$ & $T_{\mathrm{10 \, mbar}}$ (K) & $MMW$ \\
\hline
\multirow{2}{*}{Blc-EQ-S} & CO$_2$+CH$_4$ & 1.17 & 1449.47 & $-3.94_{-4.20}^{+1.68}$ (2.00$\sigma$) & $-1.73_{-0.73}^{+0.66}$ (4.62$\sigma$) & $138_{-39}^{+48}$ & $2.64_{-0.28}^{+1.16}$ \\
 & All & 1.25 & 1448.35 & $-3.63_{-5.13}^{+1.75}$ (N) & $-0.96_{-0.57}^{+0.45}$ (4.16$\sigma$) & $161_{-54}^{+90}$ & 4.32$_{-1.45}^{+3.02}$ \\
\hline
\multirow{2}{*}{\shortstack{Blc-EQ-S\\Err $\times$ 1.08}} & CO$_2$+CH$_4$ & 1.00 & 1450.17 & $-4.39_{-4.82}^{+1.97}$ (1.86$\sigma$) & $-1.77_{-0.81}^{+0.66}$ (4.24$\sigma$) & $131_{-41}^{+49}$ & $2.61_{-0.26}^{+1.03}$ \\
 & All & 1.08 & 1448.90 & $-4.47_{-5.48}^{+2.42}$ (N) & $-0.89_{-0.57}^{+0.43}$ (3.75$\sigma$) & $163_{-57}^{+96}$ & $4.59_{-1.64}^{+3.26}$ \\
\hline \hline
\multirow{2}{*}{Bts-PhQ} & CO$_2$+CH$_4$+DMS$\mathrm{^{a}}$ & 1.50 & 1443.52 & $-5.54_{-6.07}^{+4.45}$ (3.89$\sigma$) & $-1.57_{-0.89}^{+0.67}$ (5.94$\sigma$) & $134_{-37}^{+80}$ & $2.77_{-0.41}^{+5.31}$ \\
 & All & 1.58 & 1444.07 & $-1.18_{-1.60}^{+0.32}$ (2.33$\sigma$) & $-0.64_{-0.44}^{+0.33}$ (3.83$\sigma$) & $328_{-142}^{+184}$ & $10.14_{-3.39}^{+3.21}$ \\
\hline
\multirow{2}{*}{\shortstack{Bts-PhQ\\Err $\times$ 1.225}} & CO$_2$+CH$_4$+DMS$\mathrm{^{b}}$ & 1.00 & 1449.56 & $-5.80_{-5.63}^{+4.28}$ (2.32$\sigma$) & $-1.83_{-1.08}^{+0.79}$ (4.54$\sigma$) & $129_{-38}^{+71}$ & $2.58_{-0.25}^{+2.91}$ \\
 & All & 1.06 & 1449.81 & $-1.62_{-5.76}^{+0.70}$ (N) & $-0.64_{-0.58}^{+0.36}$ (3.12$\sigma$) & $334_{-154}^{+189}$ & $9.71_{-4.49}^{+3.98}$ \\
\hline
\end{tabular}
\tablefoot{Testing the impact of inflated spectral error bars, potentially accounting for systematic errors or discrepancies with atmospheric models. $^{(a)}$ {$\log{(\mathrm{DMS})}=-4.60_{-1.70}^{+0.95}$ (2.13$\sigma$)}; $^{(b)}$ {$\log{(\mathrm{DMS})}=-4.86_{-2.91}^{+1.05}$ (2.03$\sigma$)}. }
\end{table*}

\clearpage
\twocolumn

\clearpage
\onecolumn

\section{Comparison between retrieval configurations}
\begin{figure}[!ht]
    \centering
    \includegraphics[width=\textwidth]{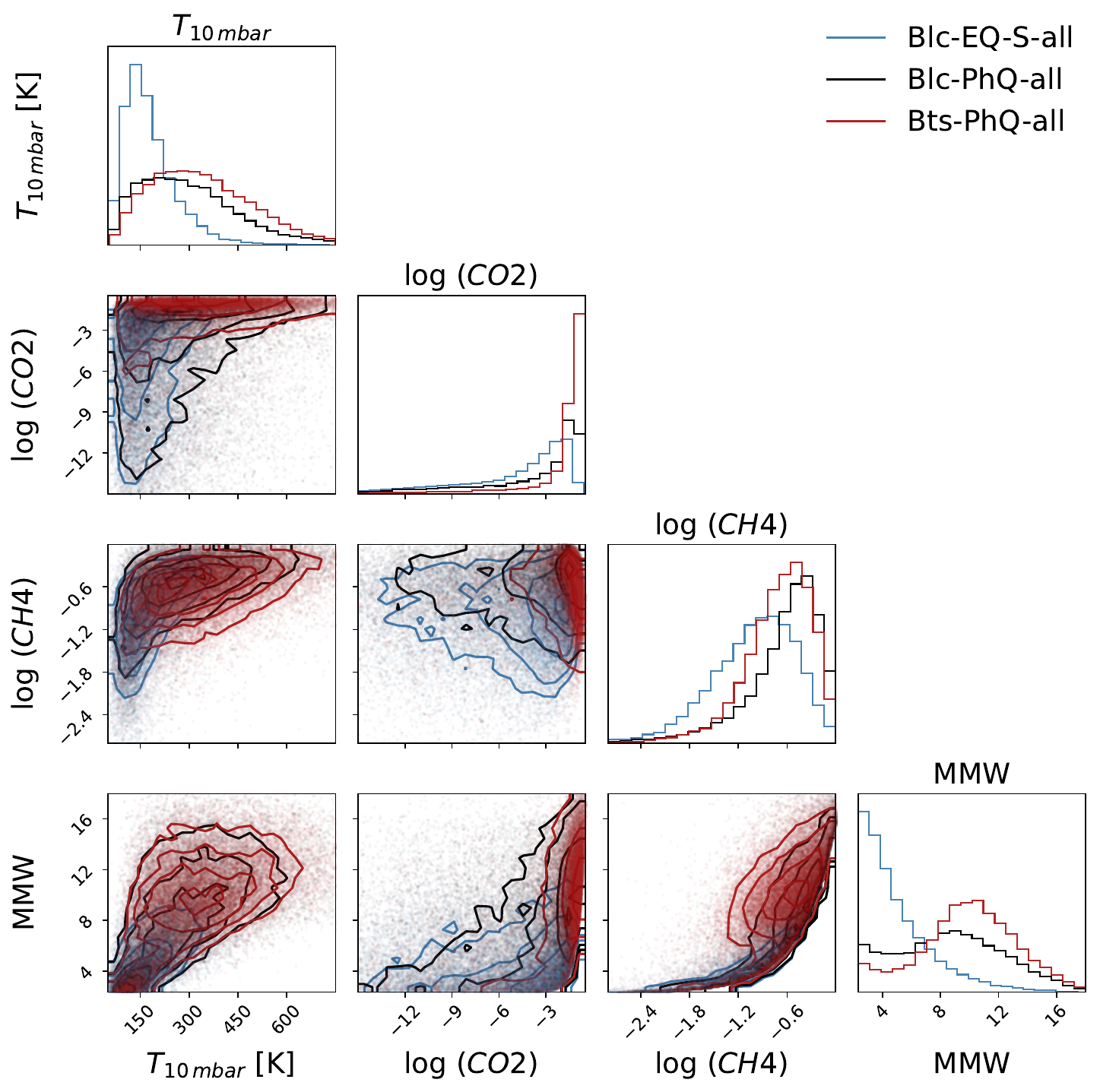}
    \caption{Posterior distributions of some selected parameters representative of the major reduction families with comprehensive retrievals. Blue shows our fiducial reduction corresponding to binned before fitting light curves - empirical limb darkening - with spot correction, black shows the binned before fitting light curves - \texttt{PHOENIX} model fixed quadratic limb darkening (no spot correction), and red the binned transmissions spectrum - \texttt{PHOENIX} model fixed quadratic limb darkening (no spot correction).}
    \label{fig:multi_posteriors}
\end{figure}

Figure \ref{fig:multi_posteriors} compares the posterior distributions of some selected parameters for three reduction configurations.

\clearpage
\twocolumn
\end{appendix}

\end{document}